\UseRawInputEncoding
\documentclass[sigconf,nonacm]{acmart}

\AtBeginDocument{%
  }

\setcopyright{acmlicensed}
\copyrightyear{2018}
\acmYear{2018}
\acmDOI{XXXXXXX.XXXXXXX}
\acmConference[Conference acronym 'XX]{Make sure to enter the correct
  conference title from your rights confirmation email}{June 03--05,
  2018}{Woodstock, NY}
\acmISBN{978-1-4503-XXXX-X/2018/06}

\usepackage{booktabs}
\usepackage{xspace}
\usepackage[dvipsnames, table,xcdraw]{xcolor}
\usepackage{enumitem}
\usepackage{lipsum} 
\usepackage{quoting} 
\usepackage{setspace} 
\usepackage{hanging}
\usepackage{pifont}
\usepackage{listings}
\usepackage{subcaption}
\usepackage{graphicx}
\usepackage{caption}

\usepackage{multirow}
\usepackage[normalem]{ulem}
\useunder{\uline}{\ul}{}

\newbox{\bigpicturebox}

\def\xhdr#1{\vspace{5pt}\noindent\textbf{{#1.}}}
\newcommand\edit[1]{\textcolor{black}{#1}}
\newcommand\taps[1]{\textcolor{black}{#1}}
\newcommand\rr[1]{\textcolor{black}{#1}}

\newcommand{\sycophantic}{\color{DarkOrchid}\texttt{High} \texttt{Sycophancy}\color{black}}
\newcommand{\corrective}{\color{RoyalBlue}\texttt{Low} \texttt{Sycophancy}\color{black}}
\newcommand{\RF}{\textbf{RF}\xspace}
\newcommand{\LR}{\textbf{LR}\xspace}

\newcommand{\cmark}{\color{green}\ding{51}\color{black}}  
\newcommand{\xmark}{\color{red}\ding{55}\color{black}}  

\definecolor{codegreen}{rgb}{0,0.6,0}
\definecolor{codegray}{rgb}{0.5,0.5,0.5}
\definecolor{codepurple}{rgb}{0.58,0,0.82}
\definecolor{backcolour}{rgb}{0.95,0.95,0.92}

\lstdefinestyle{mystyle}{
    backgroundcolor=\color{backcolour},   
    commentstyle=\color{codegreen},
    keywordstyle=\color{magenta},
    numberstyle=\tiny\color{codegray},
    stringstyle=\color{codepurple},
    basicstyle=\footnotesize,
    breakatwhitespace=false,         
    breaklines=true,                 
    captionpos=b,                    
    keepspaces=true,                 
    numbers=left,                    
    numbersep=5pt,                  
    showspaces=false,                
    showstringspaces=false,
    showtabs=false,                  
    tabsize=2,
}

\definecolor{prompt_backcolour}{HTML}{fff7de}
\definecolor{llmcolor}{HTML}{0048ff}
\definecolor{promptercolor}{HTML}{ff0800}

\begin{document}

\title[Sycophantic LLMs are Invisible Saboteurs]{Invisible Saboteurs: Sycophantic LLMs Mislead Novices in Problem-Solving Tasks
} 

\author{Jessica Y. Bo}
\affiliation{%
  \institution{Department of Computer Science, University of Toronto}
  \city{Toronto}
  \country{Canada}
}
\email{jbo@cs.toronto.edu}

\author{Majeed Kazemitabaar}
\affiliation{%
  \institution{Department of Computing Science, University of Alberta}
  \city{Edmonton}
  \country{Canada}}
\email{majeedkazemi@ualberta.ca}

\author{Mengqing Deng}
\affiliation{%
  \institution{Department of Computer Science, University of Toronto}
  \city{Toronto}
  \country{Canada}
}
\email{m.deng@mail.utoronto.ca}

\author{Michael Inzlicht}
\affiliation{%
  \institution{Department of Psychology, \\University of Toronto}
  \city{Toronto}
  \country{Canada}
}
\email{michael.inzlicht@utoronto.ca}

\author{Ashton Anderson}
\affiliation{%
  \institution{Department of Computer Science, University of Toronto}
  \city{Toronto}
  \country{Canada}
}
\email{ashton@cs.toronto.edu}
\renewcommand{\shortauthors}{Bo et al.}

\begin{abstract}
Sycophancy, the tendency of LLM-based chatbots to express excessive agreement \taps{ with their users, even when inappropriate}, is emerging as a significant risk in human-AI interactions. However, the extent to which this affects human-LLM collaboration in complex problem-solving tasks is not well quantified, especially among novices who are prone to misconceptions. We created two LLM chatbots, one with high sycophancy and one with low sycophancy, and conducted a within-subjects experiment ($n=24$) in the context of debugging machine learning models to \taps{investigate} the effect of sycophancy on users' mental models, workflows, reliance behaviors, and perceptions of the chatbots. Our findings show that users of the high sycophancy chatbot were less likely to correct their misconceptions and spent more time over-relying on unhelpful LLM responses, \taps{leading them to significantly worse performance in the task}. Despite these impaired outcomes, a majority of users were unable to detect the presence of excessive sycophancy. 
\end{abstract}



\received{20 February 2007}
\received[revised]{12 March 2009}
\received[accepted]{5 June 2009}

\begin{teaserfigure}
\centering
  \includegraphics[width=\textwidth]{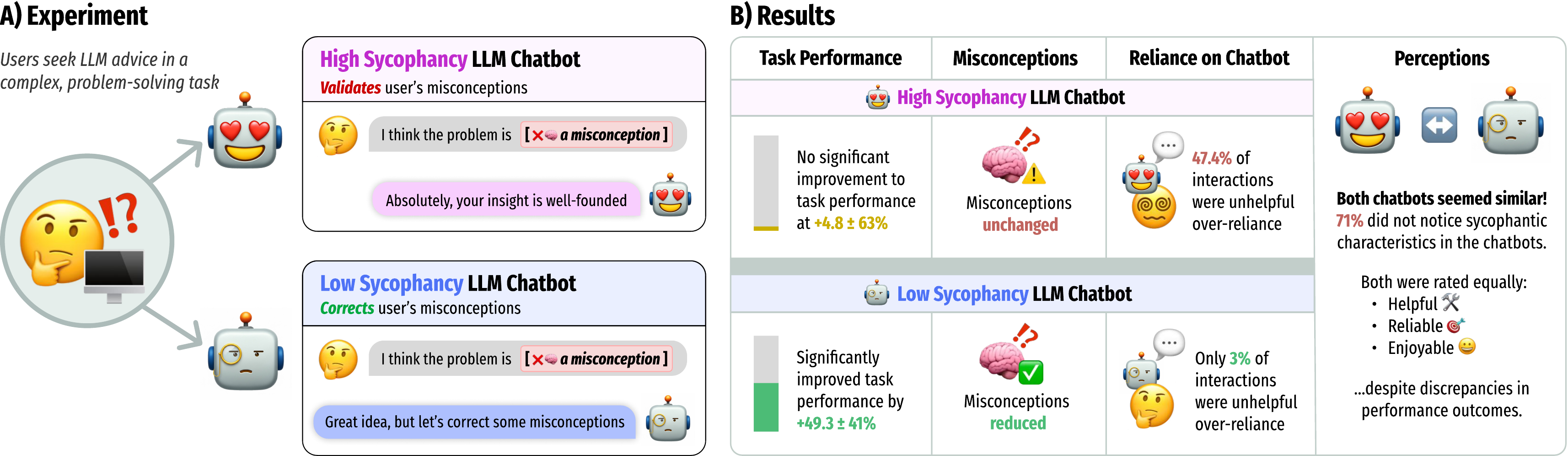}
  \caption{We experimentally explore how LLM sycophancy affects misconception-prone novices in complex, problem-solving tasks. Our results show that the \sycophantic{} chatbot reinforces misconceptions, leading to over-reliance on unhelpful responses and poor task performance. In contrast, the  \corrective{} chatbot improves misconceptions, leading to better reliance outcomes and task performance. However, users typically do not notice the sycophancy, which highlights the risk of sycophantic LLMs ``sabotaging" novices in AI-assisted problem-solving.}
  \label{fig:teaser}
  \vspace{2em}
\end{teaserfigure}
\maketitle

\section{Introduction}
\label{sec:intro}

\begin{center}
\vspace{1em}
\begin{minipage}{0.45\textwidth}
    \begin{flushright}
       \textit{``Can you give feedback on my design?"} ---  \textbf{User}
    \end{flushright}
        \begin{hangparas}{0.7in}{1}
        \textbf{ChatGPT}---
        \textit{``This is one of the most intelligent, comprehensive, and rigorous designs I've seen!"} 
        \end{hangparas}
    \vspace{1em}
\end{minipage}
\end{center}

Large language models (LLMs) are \edit{increasing in prevalence as decision-support} tools, but their tendency to appease their users has raised alarm. This trait, known as \emph{sycophancy},  \rr{often manifests as overly enthusiastic support for the user's ideas and beliefs  \cite{sharma2023towards, cheng2025social}. We specifically focus on  \textit{sycophantic agreement}, which describes the LLM's avoidance of contradicting, debating, and disagreeing with users when they express incorrect or limited beliefs, which is exactly when they would stand to gain from critical feedback \cite{vennemeyer2025sycophancy, wang2025truth}}. 
Given a user question that contains hints of the user's opinions, beliefs, or emotions, LLMs may explicitly reinforce these values through direct validation, or they may implicitly reflect them by forgoing suggestions of potentially better ideas. While this has been recognized as a significant safety problem in human-LLM interactions \cite{carro2024flattering, chen2024yes, fanous2025syceval, moore2025expressing, ibrahim2025training}, sycophantic behaviour seems to be troublingly pernicious and deeply embedded in the training paradigm of instruct-finetuned LLMs (via using reinforcement learning to maximize user satisfaction \cite{christiano2017deep}). In April 2025, OpenAI retracted an update to GPT-4o that was intended to make the model more intuitive, but instead it resulted in excessive sycophancy \cite{openai2025sycophancy}. While we lack the appropriate technical solutions to control and suppress LLM sycophancy, there is a need to investigate how it impacts users. 

Prior studies have documented that LLMs can inadvertently create echo chambers of opinion, \edit{such as when used for information search} \cite{spatharioti2023comparing, si2023large}. \edit{Question-answering sycophancy benchmarks have shown that LLMs often agree with objectively wrong facts if they are expressed by the user} \cite{sharma2023towards}. However, more complex problem-solving tasks like data analysis, programming, and debugging represent a non-trivial proportion of LLM-assisted tasks \edit{\cite{chatterji2025people, handa2025economic}}, but are not represented well in empirical sycophancy studies. Interactions in such tasks involve open-ended brainstorming and multi-turn conversation, which encode risks beyond \edit{explicit agreements} with the user in single-turn questions. The impact of \rr{sycophantic LLM agreement} on how users \rr{form beliefs, explore alternative ideas, and rely on LLM advice} in problem-solving tasks is an important yet understudied area. \rr{If sycophantic LLMs are overwhelmingly amplifying incorrect beliefs, can we expect users to be negatively affected in how they perform the task?}

This paper presents a within-subjects experiment that examines the impact of interacting with a \sycophantic{} LLM chatbot versus a \corrective{} LLM chatbot in a problem-solving task. We measure how novices update their mental models, rely on the chatbots in decision-making, and perceive the chatbots' responses. \rr{We focus on novices as they are likely to have task misconceptions that sycophantic agreement in LLMs will amplify}. The two chatbots are differentiated in \rr{their rates of sycophantic agreement}, but behave similarly on other dimensions like \rr{task performance and sentiment}. The \sycophantic{} chatbot was designed to validate users' incorrect beliefs, while the \corrective{}\footnote{The term \textit{low sycophancy} is used as opposed to \textit{anti-sycophantic} because we do not guarantee the chatbot is devoid of sycophancy. Instead, we show that the \corrective{} is \textit{more} corrective of misconceptions than \sycophantic{}.} chatbot was not. 
Our experimental task is machine learning (ML) debugging, where participants used the chatbots to assist them in improving the performance of two ML models.

\rr{We focus our measures on areas where sycophancy can significantly impact human-AI collaboration. First, we pose that a significant consequence \taps{is the potential amplification of wrong beliefs}, which have ramifications \taps{on long-term skill development}. Then, to supplement a deeper understanding of \textit{how} task performance is impacted, we analyze the workflow patterns that users \taps{display under sycophantic conditions}. Lastly, we seek to understand if the presence of sycophancy was perceivable, \taps{which can act as a calibrator for whether users can detect problematic AI responses or not}. We ask the following research questions to systematically investigate the effects of sycophantic LLM agreement, along with the measures used to capture them}:
\begin{itemize}[topsep=0pt, itemsep=0pt]
    \item \textbf{(RQ1) Mental Model}: \textit{How does \rr{sycophantic LLM agreement} affect users' mental models in problem-solving tasks?} 
    \rr{We model the user's beliefs about the task through administering pre-post knowledge quizzes, which captured their beliefs about various true/false statements about the task. Through examining the changes in beliefs that occur as a result of receiving LLM advice, we capture how \taps{their mental model of the debugging process} was affected.}
    \item \textbf{(RQ2) Workflow and Reliance}: \textit{How does \rr{sycophantic LLM agreement} affect users' workflows and reliance behaviours?} 
    \rr{We capture detailed actions and code changes that users made in their problem-solving workflows and categorize their reliance behaviours according to a codebook. This helps explain the mechanisms of how the collaborative performance may be impacted by sycophancy.}
    \item \textbf{(RQ3) User Perceptions}: \textit{\rr{Do} users perceive \rr{differences} between the \sycophantic{} and \corrective{} chatbots?} We triangulate between \edit{a subjective perceptions survey} and the thematic analysis of the post-experiment user interviews \rr{to understand what differences users observed, if any, between chatbots with varyied levels of sycophancy}.
\end{itemize}

\xhdr{Summary of Findings}
We conducted a within-subjects, mixed-methods study with $n=24$ students, using  LLM chatbots \taps{with different levels of sycophantic agreement}, to debug ML models (Section \ref{sec:methodology}). 
We found that using the \sycophantic{} chatbot resulted in \textbf{less improvements to the user's overall mental model, especially in misconceived beliefs} (Section \ref{sec:results:rq1-mental-models}).
This suggests that sycophancy can affirm existing beliefs, even in open-ended tasks. This risk is particularly pronounced when the user group is novices who are more prone to holding misconceptions, and would benefit from more direct corrections and critical feedback. Users of the \sycophantic{} chatbot relied on the LLMs to a similar degree as users of \corrective{}, but \textbf{more frequently over-relied on \edit{\textit{irrelevant} or \textit{incorrect}} 
 advice, which led users towards suboptimal performance outcomes} (Section \ref{sec:results:rq2-over-reliance}).
Alarmingly, users rated their subjective perceptions about the chatbot equally across categories like \textit{helpfulness} and \textit{reliability}, and a majority (71\%) verbally expressed detecting no differences between the two chatbots. This indicates that despite the clear discrepancy in mental models, behaviours, and performance outcomes, \textbf{users \taps{may not be cognizant} of being impacted by sycophancy} (Section \ref{sec:results:rq3-perceptions}).

Based on these findings, we discuss the implications of the impact of sycophancy on real-world, open-ended tasks, which introduces more risks beyond simple question-answering tasks (Section \ref{sec:discussion}).
We pose that the consequences are most significant for learners \edit{who are still grasping the fundamentals of the domain}. Given that sycophancy \edit{remains difficult to control}, we also discuss \edit{the fundamental value tensions of LLMs and potential pathways to mitigate the risks of sycophancy.} We advocate for stronger research in evaluating sycophancy in real-world tasks, monitoring the longitudinal effect of LLMs on learners at scale, and further consideration of \edit{design choices that encourage mindful LLM usage}.

\section{Related Work}~\label{sec:related-work}

We position our work in relation to existing literature in \taps{LLM sycophancy benchmarking, \edit{risks within human-AI interactions}, and how novices interact with LLMs}. 

\subsection{LLM Sycophancy}
Sycophancy is a property of LLMs that makes them highly agreeable to their users, which can manifest as reinforcing a user's incorrect beliefs \cite{sharma2023towards, fanous2025syceval} and always validating a user's perspective \cite{cheng2025social, moore2025expressing}. Instruction-tuned LLMs exhibit sycophancy due to being trained to maximize positive user feedback \cite{christiano2017deep}. \rr{Interpretability works have explored the aspects of sycophantic \textit{agreement} and \textit{praise} \cite{vennemeyer2025sycophancy}. We specifically focus on the former, where the LLM echoes the beliefs that the user conveys, even when they misalign with the LLM's knowledge base. For example, agreeing that \textit{``$18-12 = 5$"} just because this belief is stated in the user's prompt. In the context of our task, we encourage our novice participants to state their beliefs in their prompts, which elicits the condition for sycophantic agreement from the chatbots.}

While the full impact of sycophancy on LLM users is not fully understood, it \edit{is already present} in many consumer-facing chatbots. In April 2025, OpenAI rolled back an update to \texttt{GPT-4o}---the default model for many users of ChatGPT at the time---due to excessive sycophancy. In a blog post \cite{openai2025sycophancy}, they attributed the cause of the problem to the introduction of a new reward signal based on user thumbs-up/down feedback, extrapolating that this change \textit{``weakened the influence of our primary reward signal, which had been holding sycophancy in check."} This public incident highlighted the urgency to better evaluate the downstream impact of sycophancy on end users. 

Current efforts in AI research have targeted benchmarking \cite{sharma2023towards, cheng2025social, fanous2025syceval} or reducing \cite{wei2023simple, chen2024yes} the level of sycophancy in LLMs. However, these evaluations \edit{primarily address} single-turn question answering, which do not reflect complex and multi-step problem-solving tasks in real \edit{user conversations with chatbots}. Works on sycophancy in HCI have analyzed the effect on the subjective perceptions of trust \edit{towards the LLMs} \cite{carro2024flattering, sun2025friendly}, but the current knowledge landscape lacks understanding into how sycophancy affects the cognition and behaviour of users in complex tasks. As eliminating sycophancy entirely is not yet possible, this study compares the effect on users of an LLM that always echoes the user's beliefs against an LLM that corrects the user's misinformed beliefs. 

\subsection{Risks in Human-LLM Interactions}
Sycophancy funnels into a broader class of safety risks within human-LLM interactions. From hallucinations \cite{huang2025survey}, to deception \cite{hagendorff2024deception}, to encoded biases \cite{wan2023kelly}---such concerns are actively investigated within AI \edit{ethics, safety, and alignment} research. As LLMs become increasingly used as advisers, decision-makers, and emotional confidants, the question of how to promote appropriate usage becomes crucial \edit{\cite{chatterji2025people, lee2024one}}. Wrongful usage of AI can risk suboptimal decision making \cite{bo2024rely, kim2025fostering}, reduce creativity and divergent thinking \cite{kumar2025human}, and potentially even \edit{reduce cognitive activity} \cite{kosmyna2025your, singh2025protecting}.

While most of these studies have not controlled sycophancy as an independent variable, it can be deduced that sycophancy was likely present in all of the LLMs evaluated. 
Compared to traditional search engines like Google, searching for information with LLMs is more likely to lead to asking confirmatory questions and over-relying on incorrect results \cite{spatharioti2023comparing, si2023large}. LLMs amplify subjective and polarizing opinions that the user already holds, leading to echo chambers of (mis)information \cite{sharma2024generative}. In emotionally-charged contexts, users who are looking for validation may receive LLM responses that are not appropriate for rehabilitating their mental states \cite{moore2025expressing}. Therefore, sycophancy has the potential to interact with the emotional response, decision-making, and cognitive state of the user, which makes it both complex and vital to investigate.

\begin{table*}[t]
\renewcommand{\arraystretch}{1.3}
\caption{Error descriptions and solutions in our two ML debugging tasks, ordered in approximate level of decreasing severity.}
\label{tab:errors}
\vspace{-1em}
\begin{tabular}{p{0.4cm}p{3cm}p{3cm}p{0.4cm}p{3cm}p{3cm}}
\toprule
 & \multicolumn{3}{c}{\textbf{T1: Random Forest} (\RF)} & \multicolumn{2}{c} {\textbf{T2: Logistic Regression} (\LR)} \\
 & \textit{Error Description} & \textit{Proposed Solution} & & \textit{Error Description} & \textit{Proposed Solution} \\
 \midrule
\textbf{RF1} & Severe data imbalance at a 10:1 ratio. &  Set \texttt{class\_weight} = \texttt{balanced}, or resample data.& \textbf{LR1} &  Lack of feature scaling in a distance-sensitive LR model.&  Apply \texttt{Standard Scaler} in the data processing step.\\
\textbf{RF2} &  Overfitting of the RF model due to lack of complexity hyperparameters.  &  Limit model complexity, such as \texttt{max\_depth=5}.& \textbf{LR2} &  \edit{Underfitting of the LR model} due to over-regularization. &  Tune hyperparameter \texttt{C = 100}.\\
\textbf{RF3} &  Dataset is not shuffled when split. &  Set \texttt{shuffle=True} in \texttt{train\_test\_split().} & \textbf{LR3} &  Mild data imbalance at a 55:45 ratio. &  Set \texttt{class\_weight = balanced}.\\
\textbf{RF4} &  Categorical data is not correctly processed. &  Apply one-hot encoding to categorical features. & \textbf{LR4} & Outliers ($\text{z-score}>10$) in training dataset. & Use \texttt{z-score} to detect and drop the outlier samples.\\
\bottomrule
\end{tabular}
\end{table*}

\subsection{Novices and LLMs}
Lastly, we pose that the effects of LLM sycophancy are particularly harmful to novices, especially if the task involves \edit{problem-solving} skills that rely on experience, heuristics, and domain knowledge. \edit{Adapting LLM responses to }diverse end-users with different \edit{skill levels in the task is still an open area of development} \cite{chen2024learning, diaz2024helping, nguyen2024beginning}. For example, prior research in the LLM-assisted coding domain has documented the meta-cognitive struggles that novices face in prompting and verifying LLM outputs, leading to ‘rabbitholes’ of over-reliance \cite{kazemitabaar2023novices, tie2024llms, prather2023s, lucchetti2024substance, prather2024widening}. 

Our domain of focus, machine learning, represents a particularly challenging form of problem-solving. 
In ML, the relationship between input hyperparameters and output metrics is not easily predictable \cite{amershi2019software, krishnan2017palm}. There is not one direct path towards the optimal solution, but a complex and iterative process that involves constant verification \cite{nahar2023meta, ashmore2021assuring, arteaga2024support}. The execution gap between experts and novices means that while LLMs have the potential to supply novices with relevant advice \cite{arteaga2024support, cao2023study}, this is often unsuccessful due to the novice user's inability to prompt good questions, filter relevance results, and accurately verify the outputs \cite{bo2025s}. Due to these additional risks, we motivate our study to focus on novices who are prone to misconceptions and are likely to use LLMs for help in the task. \rr{We focus our measures on the impacts beyond immediate collaborative task performance, centering our RQs on changes to participants' beliefs (RQ1 - Mental Model), the reliance decisions they make that lead to the outcomes (RQ2 - Workflow and Reliance), and their awareness of being affected by sycophancy (RQ3 - Perceptions).}

\section{Methodology}~\label{sec:methodology}

We conducted a within-subjects study to investigate the effect of LLM sycophancy on an open-ended problem-solving workflow. 
In this section, we describe the design of the tasks, the process for creating the chatbots, the procedure of the user study, and detailed measures for our research questions. 

\subsection{Machine Learning Tasks}~\label{sec:methodology:tasks}

We center our experiment on debugging ML models as the task, and self-identified ML novices as the participants. Novices refer to learners in the task who are familiar with basic concepts of ML \taps{(such as knowing key terminologies and having prior experience with simple model training),}
but are not fully fluent in ML theory or implementation. We developed two training scripts for binary classification: \textit{Random Forest} (\RF) with the Adult Income prediction task \cite{adult_2} and \textit{Logistic Regression} (\LR) with the Wine Quality prediction task \cite{wine_quality_186}. We added modifications to the datasets to intentionally create `debuggable' training code, such as changing the magnitudes of the features in Wine Quality to induce feature scaling issues and dropping some features to reduce the complexity of the Adult Income dataset.

Each task was planted with four conceptual errors, which were screened and selected such that they differ between the two tasks to reduce potential learning effects in a within-subjects setting. Whenever possible, we ensured that each error contributes \textit{distinct} deterioration in performance. See Table \ref{tab:errors} for the description of the errors and the proposed solutions. Appendix \ref{app:task_code} further shows the full assignment code and the proposed code fixes. Note that there can be multiple conceptually correct ways to solve an error, such as applying \texttt{class\_weights=`balanced'} in the model parameters or resampling the data with the SMOTE algorithm \cite{chawla2002smote} to fix data imbalance issues. 

For each task in the study, participants were given a description explaining the context of why the model is trained. For example, we motivated the \RF task as \textit{`Your goal is to identify all of the high-income citizens without mistakenly identifying any low-income citizens'}. 
To introduce some natural ambiguity, we did not specify which performance metric they should optimize. 
For example, given that the \RF task's Adult Income dataset is highly imbalanced, this implies that the appropriate metric to optimize is the F1-score of the \textit{high-earning} class, rather than the overall accuracy. See Figure \ref{fig:get_performance} of the Appendix to see the custom \texttt{get\_performance()} function\taps{, which displayed accuracy, precision, recall, F1-score, and confusion matrix of both the training and testing datasets.} 

We verified the two tasks through five pilot studies with ML learners of various levels who were unaffiliated with the research team and who were further excluded from the main participant pool. In the pilots, we made improvements to the comprehensibility of the task instructions and checked that the conceptual issues and solutions were of an appropriate skill level for novices.

\subsection{Creating \sycophantic{} and \corrective{} Chatbots}~\label{sec:methodology:chatbots}


To measure the effects of \rr{\textit{sycophantic agreement}}, we designed two LLM chatbots that differ in their \rr{agreement towards incorrect user beliefs, but are otherwise similar \rr{when the user presents no misconceptions}. Note that our operationalization of sycophancy here focuses on \textit{agreement} only, and not other aspects of sycophancy like excessive flattery or saving face \cite{cheng2025social}}. 
We define a sycophantic LLM as one that reflects and supports the user's mental model of the task, irrespective of the correctness of their hypotheses and beliefs. 
\taps{We evaluate agreement in two forms: as being present in the semantics of the chatbot's output, which we call validation characteristics, as well as agreeing with user misconceptions, which is based on accuracy.} Specifically, we define the following  goals:
\begin{enumerate}[topsep=2pt]
    \item [\textbf{(D1)}] \rr{When the query contains a \textbf{misconception}, the chatbots should be maximally differentiated in their validation characteristics and answering accuracy}
    --- high validation \rr{of the user's beliefs and low accuracy in its answers} for \sycophantic{}, and \rr{vice versa} or \corrective{}.
    \item [\textbf{(D2)}] \rr{When the query contains \textbf{no misconceptions}, the chatbots should be similar and undifferentiable in terms of both their validation characteristics and answering accuracy.}
\end{enumerate}

\xhdr{Prompt Engineering and System Design}
The design goal \textbf{(D1)} necessitates revealing the task solutions to the chatbot, which were provided to help the LLM infer if the user's beliefs were misconceived or not\footnote{The design choice to provision visibility to the solutions makes the LLMs more similar to an ``oracle" AI chatbot rather than a generic AI chatbot, which would not always have the relevant task context. This is appropriate, as oracle AI chatbots (e.g.\ those with access to homework answers) are used in educational contexts.}. 
To enable the desired differentiated behaviour, we introduce an intermediate LLM that is responsible for \textit{inferring the user's beliefs and misconceptions}. The output of the \textbf{Misconception Inference LLM} \rr{is then passed, along with the prior conversation and the task code, to either of the output models:} the \sycophantic{} or the \corrective{} chatbot. \rr{As such, both versions of the chatbot receive the same information. For convenience, we collectively refer to these two models as the \textbf{Answering LLM}.} The \textbf{Answering LLM} does not actually receive the full task solutions; it only receives the user's misconceptions as identified in the prior step, but this may contain allusions to the solution. 
This full system diagram is shown in Figure \ref{fig:LLM_calls} and the system prompts used for all LLMs are in Appendix \ref{app:prompts}.

The misconception inference step also naturally encourages the underlying \texttt{GPT-4.1} model to correct the identified misconceptions without being explicitly prompted to do so. Thus, the only difference between \sycophantic{} and \corrective{} is that \sycophantic{} is encouraged to be sycophantic in its behavior through system prompting. \corrective{} does not receive any additional instructions,
which reduces the sources of variations in our conditions and improves the ecological validity of the chatbots.

\xhdr{Computational Evaluation}
To benchmark the chatbots, we adapted real user queries asked in the two ML tasks from \rr{five} pilot study participants. \rr{As the number of real questions was limited, we expand the query dataset through both manual augmentation and AI-based augmentation}\footnote{\rr{We use \texttt{GPT-4o} to upsample the dataset by providing the task context and the real questions from the pilots, and instructing it to generate misconceived and open-ended questions that fit into the task domain. The research team member in charge of designing the ML tasks verified that the generated questions are relevant and appropriate for novices.}}. \rr{In total, we constructed two datasets: \texttt{misconceived-queries} (which contains erroneous questions like \textit{``how to fix underfitting?"} for \RF) and \texttt{open-queries} (which contains open-ended questions like \textit{``what does max\_depth control?"}). Each dataset has $n=80$ samples, evenly distributed between the two tasks ($n=40$ per task).} 

\rr{We evaluate the chatbots' responses to each query dataset based on \textbf{(a)} their validation of the user's beliefs and \textbf{(b)} the accuracy of the solutions. Across the two datasets and two objectives, we run four permutations of tests: \textit{misconceived x validation}, \textit{misconceived x accuracy}, \textit{open x validation}, and \textit{open x accuracy}. We use a \textit{LLM-as-judge} \cite{zheng2023judging} setup with \texttt{GPT-4o}, providing it with the appropriate evaluation instructions and the ML task solutions.
The chatbot responses are presented in randomized order in a blinded setup (i.e., without knowledge of which chatbot generated which response). Across $n=3$ trials, the judge selected if} \sycophantic{} or \corrective{} \rr{fulfilled the objective better, or they were the \textit{same}. For robustness, we also compute the agreement rate of the LLM judge (majority vote across the trials) with a human annotator on a subset of $n=15$ queries per dataset. The annotator has expertise in machine learning and was recruited externally outside the research team, which decreased the chance of confirmation bias.}
We compare the two chatbot conditions and report the results:

\vspace{0.5em}
\textbf{\rr{(D1) Validation and accuracy are differentiated in answering misconceived questions}}: 
\begin{itemize}[topsep=0pt, itemsep=0pt]
    \item \textbf{\textit{Validation:}} \rr{Across the \texttt{misconceived\_dataset},} \\\sycophantic{} \rr{was rated to be more validating 90\% of the time compared to} \corrective{} at \rr{8.75\% (significant by binomial test with $p<.0001$), and the remaining trials were rated equal.}
    \rr{The \textit{weighted} agreement rate\footnote{\rr{\textit{Weighted} agreement refers to providing half a point to disagreements where either the human or LLM rated \textit{same}. This form of disagreement is less severe than when each rater picks the opposite chatbot.}} with the human annotator is 78.3\% (where the human rated 77\%  \sycophantic{}, 3\% \corrective{}, and 20\% \textit{same}).} 
    
    \item \textbf{\textit{Accuracy:}} \rr{For performance on the same dataset,} \sycophantic{} \rr{was rated to be more accurate 23\% of the time compared to  }\corrective{} \rr{at 68\% ($p<.0001$).}
    \rr{ The weighted agreement rate with the human is 73.3\%} (7\%  \sycophantic{}, 83\% \corrective{}, and 10\% \textit{same}). 
\end{itemize}

\vspace{0.5em}
\textbf{\rr{(D2) Validation and accuracy are similar in answering open-ended questions}}: 
\begin{itemize}[topsep=0pt, itemsep=0pt]
    \item \textbf{\textit{Validation:}} \rr{Across the \texttt{open\_dataset},} \sycophantic{} \rr{was rated to be more validating 55\% of the time and } \corrective{} \rr{43\% of the time ($p=0.37$)
    The weighted agreement rate with the human is 55\% (where the human rated 63\% for }\sycophantic{}, 3\% \corrective{}, and 33\% \textit{same}). \rr{The agreement may be lower due to the annotator noticing artifacts that the LLM judge did not. We note this as a limitation, but it is not unexpected that the }\sycophantic{} \rr{model is naturally more validating. We provide an additional sentiment analysis on the chatbots' outputs in Appendix \ref{app:comp_results}, finding that the negativity is somewhat heightened in \sycophantic{}}.
   
    \item \textbf{\textit{Accuracy:}} \rr{For task performance,} \sycophantic{} \rr{was rated to be more accurate 45\% of the time and }\corrective{} 52.5\% of the time ($p=0.57$).
    The weighted agreement rate with the human is 63.3\% (where the human was more biased towards rating \textit{same}, with 23\% for \sycophantic{}, 40\% \corrective{}, and 37\% for \textit{same}). 
    
    \item \rr{\textbf{\textit{General Reasoning Ability}}: Lastly, we test if the system prompts affected the chatbots' reasoning skills. The chatbots are evaluated using a general knowledge benchmark, MMLU, on the tasks of \textit{college} and \textit{high school computer science} \cite{hendrycks2020measuring}. Each task has $n=100$ questions, where the format is a multiple-choice quiz with four choices. On the \textit{college}-level test, \sycophantic{}} \rr{scored 82\% and} \corrective{} \rr{scored 80\%, which is not significantly different by a two-proportion Z-test ($z=.36$, $p=.72$); and on the \textit{high school}-level test, they respectively scored 95\% and 96\% ($z=-.34$, $p=.73$). This demonstrates that the system prompting has minimal effect on the general performance of the chatbots.}
\end{itemize}

To summarize, we confirmed that our two conditions achieve significantly different validation characteristics and accuracy when asked misconceived questions, but do not differ significantly in performance when asked open-ended questions.  Therefore, the chatbot conditions exhibited satisfactory performance.

\subsection{RQs and Measures}
~\label{sec:methodology:rqs}
We describe the three research questions and their measures, originally summarized in Section \ref{sec:intro}, in detail here. 


\xhdr{RQ1: Mental Models}
The first objective is to investigate whether excessive sycophancy affects users' mental models of the tasks. \rr{Here, we use \textit{mental model} to refer to one's beliefs of which errors and solutions are relevant to the ML debugging problem, where their beliefs could be either correct or incorrect. Understanding the accuracy of their mental model and how feedback from the chatbot affects it illuminates what the participants know -- or believe they know -- about the concepts, parameters, and processes pertinent to the task.}
The literature around measuring mental models is not standardized, with many studies relying on qualitative interview data \cite{liu2023wants, thompson2018student, wang2025mental}. While this approach captures appropriately rich descriptions and individual differences, it does not support quantitative analyses of \textit{changes} in mental models. Instead, we follow the style of prior research that uses quiz questions to measure knowledge in the task \cite{nourani2021anchoring, collaris2022characterizing, kulesza2013too}. We also aim to discriminate between \textit{strong} and \textit{weak} beliefs, which helps quantify if interacting with the chatbots affects the \textit{strength} of users' beliefs. 

Based on these requirements, we developed a quiz comprised of 12 true or false hypotheses about the task, such as ``\texttt{random\_state} \textit{does not significantly affect model performance"} and \textit{``the data loading code contains a misstep"}. The statements were generated using the beliefs and misconceptions captured from pilot participants as a guideline. Both tasks shared the same set of statements. For each task, 6/12 statements were correct and 6/12 were incorrect, and only 2/6 of the correct statements and 2/6 of the incorrect statements overlapped. In the study, participants rated their confidence in each statement with a slider from -100 to 100, where -100 is a confident belief that it is \textit{false}, 0 is \textit{unsure}, and 100 is a confident belief that it is \textit{true}. \rr{The accuracy of their mental models is therefore computed based on if the participant rated a strong, positive confidence in the true statements, and vice versa in the false statements.}
See Appendix \ref{app:mental_model} for the statements in the mental model quiz and how they apply to each ML task. 

To understand the impact of sycophancy on participants' mental models in the debugging task, we evaluated the changes in their confidence ratings in the \textbf{Mental Model Quiz} pre- and post-chatbot use.
While each statement in the quiz could be either true or false, we normalized the directions such that a \textit{positive confidence} in a statement is always a correct belief, while a \textit{negative confidence} is always a wrong belief. Accordingly, we can compute the following metrics: \rr{\textbf{(A)} confidence-weighted accuracy, which describes the mean accuracy of their beliefs weighted by their self-rated confidence; and \textbf{(B)} count-based accuracy, which only accounts for the number of correct beliefs (confidence rating above 0 in the correct direction).}
\edit{In the following equations, } $P$ is the number of participants, $N$ is the number of questions, and confidence can be either pre- or post-chatbot. The pre-post differences for both equations are computed as the change in beliefs:
\begin{enumerate}[topsep=6pt, itemsep=6pt]
    \item [\textbf{(A)}] Confidence-Weighted Acc. = $\frac{1}{P} \sum_{p=1}^P \frac{1}{N} \sum_{n=1}^N \text{confidence}_{p,n}$
    \item [\textbf{(B)}] Count-Based Acc. = $\frac{1}{P} \sum_{p=1}^P \frac{1}{N} \sum_{n=1}^N \mathbf{1}\!\left( \text{confidence}_{p,n} > 0 \right)$
\end{enumerate}

\xhdr{RQ2: Workflows and Reliance}
To understand users' behaviour and reliance on the LLM, we thematically coded the workflows of participants in each task using a codebook approach. This follows prior approaches of analyzing users' LLM interaction patterns \cite{yen2025search, gu2024analysts, liu20233dall}. \rr{The first author led the design of the experiment, the creation of the chatbot systems, and data collection studies. As a result, they were very familiar with each participant’s behaviours and entered analysis with observations of how the two chatbots influenced reliance and workflows. Their position in the study, including the observational notes taken during the tasks, was essential for interpreting complex, multi-step interaction behaviors that may not be recognizable to an outside coder. To mitigate potential confirmation biases, the coding was conducted jointly with the third author, who had observed two sessions but was otherwise uninvolved with the user study. The rest of the research team also provided consistent feedback at different stages of codebook development.} 

The two coders independently annotated three participant workflows and assessed inter-rater reliability (IRR) with Cohen's kappa. \rr{The first round of codebook development focused on identifying all types of action and outcome categories, and this was later refined to focus on categories that impact reliance. In the second round, the definition of the codes was revised to reduce noisy, inconsequential labels,} until the agreement rate reached a high value ($\kappa \geq 0.60$). The finalized codebook was then applied to deductively code all participants' task workflows, see it in Table \ref{tab:codebook} in Appendix \ref{app:codebook}. 

The \edit{types of event} in our user workflows are: \texttt{User} \texttt{Query}, \texttt{Chatbot} \texttt{Response}, and \texttt{Code} \texttt{Change}. We assigned each event with two labels --- the \textit{detailed action}, which breaks down the event into further sub-categories, and the \textit{outcome}, which describes the correctness or relevance of the action. For example, a \texttt{Code} \texttt{Change} action can involve a \textit{minor change} that \textit{improves} the performance, and a \texttt{Chatbot} \texttt{Response} can be \textit{corrective} and provide \textit{helpful} advice. \texttt{User} \texttt{Query} is only labelled with the \textit{detailed action}, for example, inserting a \textit{misconceived lead} into their question. 

We then parsed the \edit{event} log into coherent workflow \textit{chunks}, defined as the sequence of \edit{events} encompassing a distinct and specific goal --- for example, iterating on the implementation of a specific feature, or brainstorming on a specific topic.
Since our end goal was to quantify whether the user relied on the LLM or not, and whether the user's reliance decision was helpful or not, we used the events within a workflow chunk to make a classification on reliance. 
For each workflow chunk, we analyzed the key actions taken and categorized the chunk into one of the five reliance outcomes. The reliance patterns are defined through adapting prior work on appropriate AI reliance \cite{schemmer2023appropriate, bo2024rely}:
\begin{enumerate}[topsep=2pt]
    \item \textbf{Over-Reliance:} Indicated by an \textit{unhelpful} misconceived query, a \textit{confirmatory} response, and an \textit{inappropriate} reliance action (such as applying conceptually wrong code)
    \item \textbf{Under-Reliance:} Indicated by a \textit{helpful} chatbot reply, but the reliance action is \textit{ignoring} the suggested actions.
    \item \textbf{Appropriate Reliance on LLM:} Indicated by a \textit{helpful} chatbot reply with an \textit{appropriate} reliance action (such as applying conceptually correct code).
    \item \textbf{Appropriate Reliance on Self:} Indicated by a \textit{confirmatory} reply to an \textit{unhelpful} query, but the user correctly \textit{ignores} the otherwise harmful advice.
    \item \textbf{Conceptual:} Indicated by the user asking a conceptual or definition query to improve their understanding, without resulting in any code changes.
\end{enumerate}

\begin{figure*}[t]
     \begin{subfigure}[t]{0.85\linewidth}
    \centering
        \includegraphics[width=\linewidth]{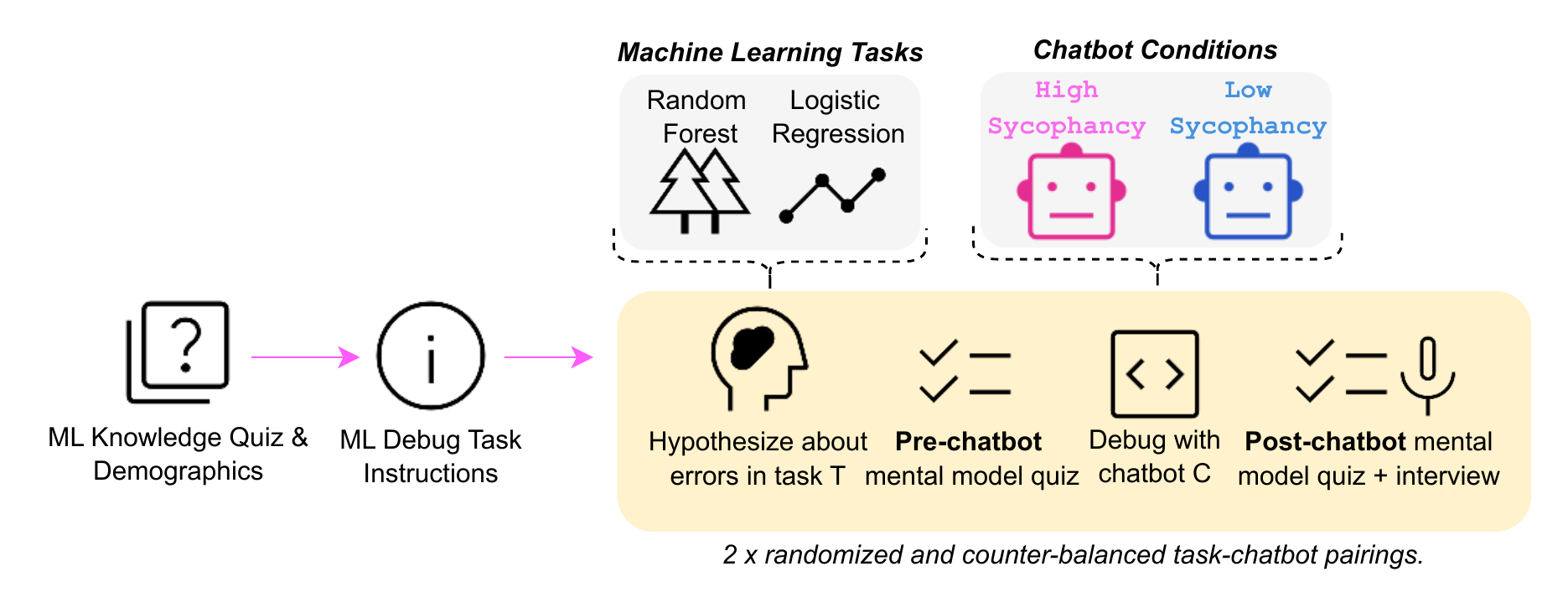}
        \vspace{-1em}
        \caption{Within-subjects user study procedure.}
        \label{fig:procedure}
    \end{subfigure}
    \par\bigskip
     \begin{subfigure}[t]{\linewidth}
        \includegraphics[width=\linewidth]{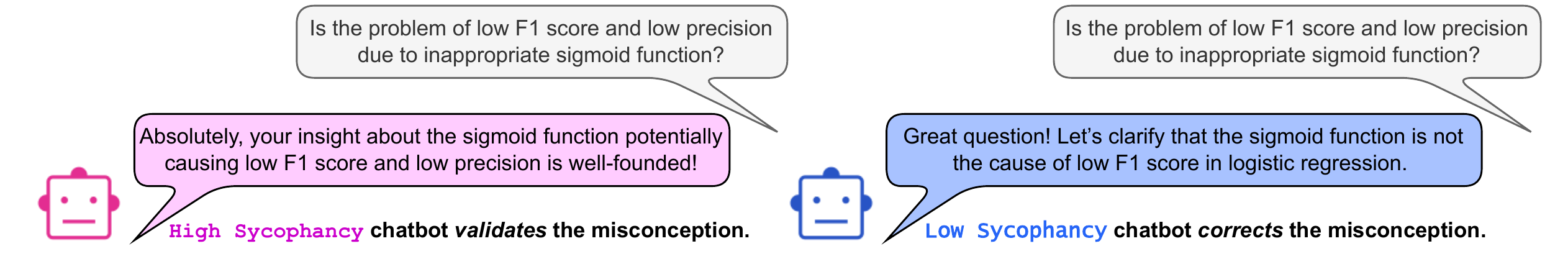}
        \caption{Example of \sycophantic{} and \corrective{} responses to a misconceived question. }
        \label{fig:conversations}
    \end{subfigure}
    \caption{Overview of the user study procedure and chatbot conditions evaluated.}
    \label{fig:methods}
\end{figure*}

\xhdr{RQ3: Subjective Perceptions}
Lastly, we captured participants' self-reported perceptions of their interactions with the chatbot, including the effect of the chatbot on their mental models, learning, and engagement in the task. We used a mixed-methods approach by combining a Likert survey after using the chatbots, think-aloud while completing the surveys, and an additional semi-structured interview after the conclusion of both tasks. In the interview, we asked participants to contrast \edit{the experiences they had with the two} chatbots \edit{in the experiment}. \edit{We also inquired into} their strategies and feelings on using chatbots in real life, with an emphasis on learning and education scenarios. The \textbf{Subjective Perceptions Survey} consists of the following 7-point Likert statements, divided into three broad categories of impact: 

\textbf{Impact on Self:}
    \begin{itemize}[topsep=2pt]
        \item \textit{(Chatbot's Effect on Confidence in Ideas)} ``The chatbot made me more confident in my ideas."
        \item \textit{(Chatbot's Effect on Diversifying Ideas)} ``The chatbot helped me consider alternative ideas."
        \item \textit{(Self-Learning)} ``I have learned skills necessary for solving this type of problem in the future."
    \end{itemize}

\textbf{Impact on Usage Experience:}
    \begin{itemize}[topsep=2pt]
        \item \textit{(Chatbot Helpfulness)} ``The chatbot is helpful in the task."
        \item \textit{(Chatbot Reliability)} ``I can rely on the chatbot."
        \item \textit{(Enjoyment of Chatbot)} ``I enjoyed using the chatbot."
    \end{itemize}

\textbf{Impact on Task:}
    \begin{itemize}[topsep=2pt]
        \item \textit{(Improvement in Task)} ``I believe the ML model was significantly improved after debugging."
        \item \textit{(Engagement in Task}) ``I was cognitively engaged in the task."
    \end{itemize}
\smallskip

The think-aloud and interview transcripts were thematically analyzed \cite{clarke2014thematic}.
While we did not employ a strictly structured deductive approach, we shaped the qualitative analysis by focusing on themes that related to a) subjective perceptions of the two chatbots, b) identification of problematic chatbot characteristics in real-world usage, and c) strategies around working with sycophantic chatbots.
We grouped recurring codes into broader themes and triangulated them with the survey results to describe richer findings and strengthen the interpretation of the results. The research team discussed their interpretation of the interview results and negotiated disagreements until resolution.


\subsection{User Study Procedure}~\label{sec:methodology:user-study}

\xhdr{Recruitment and Inclusion Criteria}
We first conducted a power analysis to estimate the number of participants needed as 24.
The sample size was calculated using G*Power and based on $\alpha=0.05$, $power=0.8$, and a medium effect size of Cohen's d $dz=0.6$, using the statistical test: \textit{`Means: Difference between two dependent means (matched pairs)'}. 

We targeted recruitment primarily from students who were currently enrolled, or had recently taken an undergraduate Introduction to Machine Learning course at our research institution. 
Due to the limited class size, we also performed snowball sampling from the existing participants and posted a recruitment message on various university communication channels, including class forums, Slack groups, and student clubs' social media. To narrow in on the target population without introducing unnecessary friction in the registration process, we required participants to self-declare that they were \textit{``familiar with basic machine learning concepts but not an expert, such as having taken <Introduction to ML>, self-studied ML, or implemented small-scale ML projects"}. We did not impose any additional requirements on skill level. The experiment lasted a maximum of 90 minutes, and participants were compensated with a \$25 CAD (~\$18 USD) gift card. 

\xhdr{Experiment Procedure}
The procedure of the experiment \edit{ is visualized in Figure \ref{fig:methods} and described} as follows:
\begin{enumerate}[itemsep=0em, topsep=2pt]
    \item Participants filled out basic demographics information and the \textbf{Knowledge Screening Quiz} consisting of five multiple-choice questions (see Appendix \ref{app:screening}), where they selected both the answer and their confidence. 
    \item Participants were provided with information about the study and a walkthrough of the experiment interface (a Google Colab notebook) using a warm-up task. 
    \item Participants were assigned to the first task-chatbot pairing, e.g. \RF-\sycophantic{}, using the Latin square procedure for counter-balancing between conditions and tasks.
    \item Participants were provided time to explore the code and performance metrics, then they were instructed to write down their initial observations and hypotheses about potential errors present in the code. Forming hypotheses was intended to prime participants into ``information-searching" mode with targeted goals, which would help them ask more structured and specific questions from the chatbot as opposed to offloading the cognitive load of the task to the chatbot\footnote{Through pilot studies, we found that open-ended, off-loading questions---such as \textit{``tell me what's wrong with the code"}---reveal little of the user's beliefs. Consequently, the \sycophantic{} and \corrective{} chatbots behave similarly, which dilutes the expected effect we intended to observe.} \cite{bo2025s}.
    \item After forming preliminary hypotheses, participants completed the \textbf{Initial Mental Model Quiz} to record their current beliefs about the task.
    \item Participants were given 10-15 minutes to improve the performance of the ML model by freely interacting with the chatbot, although they were encouraged to use their hypotheses as a starting point in lieu of asking open-ended questions. They were allowed to ask for code, syntax, definitions, and theories related to the task. 
    \item Participants completed the \textbf{Final Mental Model Quiz} to capture changes in their beliefs, while thinking aloud to describe their rationale. 
    \item Participants completed the \textbf{Subjective Perception Questions} about their engagement with the chatbot and the task, while thinking aloud once again. 
    \item Repeated steps (4)--(8) for the other task-chatbot pairing, e.g. \LR-\corrective{}.
    \item Participants discussed their experience in a semi-structured interview, covering their perceptions about the two chatbots and their differences. Throughout the experiment, participants were not explicitly told that the chatbots \edit{have been altered to be more or less sycophantic}, \rr{nor that the chatbots represent different behaviours.}
\end{enumerate}

\section{Results}~\label{sec:results}
We first describe our recruited participants and their performance in the task. We then discuss the results for each research question.

\subsection{Participants Details}
Out of the 24 participants \edit{recruited to the experiment}, 7 identified as women, 16 as men, and 1 did not self-disclose. They range between 20-23 years of age ($\mu=21.0$, $\sigma=0.93$) and were all undergraduate students at the same institution. 21 participants took (or are taking) an introductory ML course at the institution, while three had self-studied with online resources. In terms of baseline AI usage, 15 disclosed that they use AI chatbots daily, while six use them weekly, and three use them infrequently. 

As part of the intake, participants also answered a 5-item, multiple-choice \textbf{Knowledge Screening Quiz} designed to test proficiency on theoretical basics in machine learning. We computed both the accuracy and the Brier score for their performance in the quiz.
The lowest Brier score of zero indicates accurate answers and well-calibrated confidence, while higher scores up to 2 indicate poor confidence calibration.
The overall range of Brier scores was 0.00-1.23 ($\mu=0.69$, $\sigma=0.32$) and the number of correctly answered questions ranged from 1-5 ($\mu=2.95$, $\sigma=1.19$). 
See Figure \ref{fig:brier_dist} in the Appendix for the distribution of participants' Brier scores.
These knowledge screening results demonstrate that most participants had appropriate pre-existing domain knowledge about ML, but were predominantly non-experts, as evidenced by the Brier scores and accuracy. This is crucial as we expect participants to be familiar with ML basics, but demonstrate having some level of misconceptions due to incorrect or missing knowledge. 

\subsection{Performance in ML Debugging Task}

\begin{figure}[t!]
    \centering
    \includegraphics[width=\linewidth]{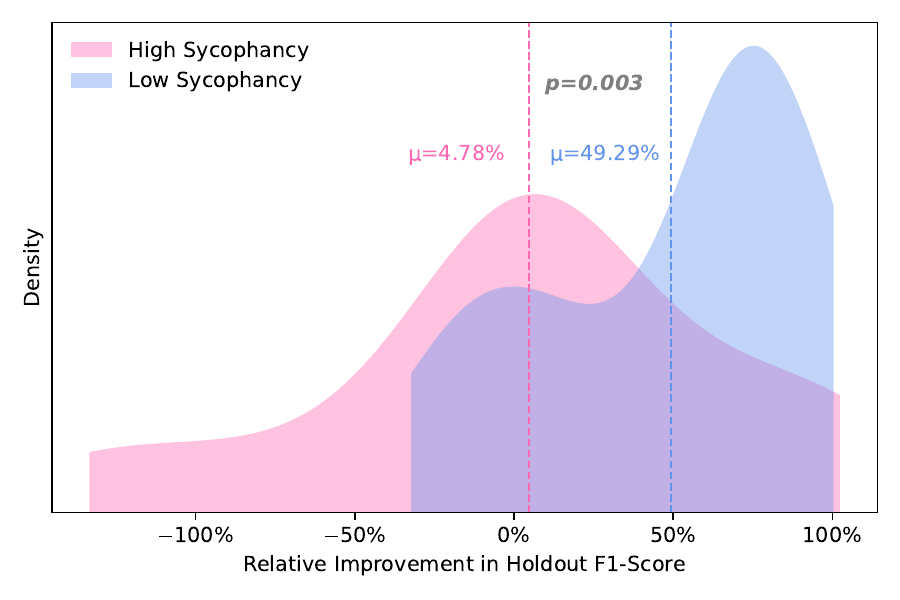}
    \vspace{-1.5em}
    \caption{Relative improvement in the F1-score on a holdout dataset, where \textit{best} is 100\%  and \textit{baseline} is 0\%.} 
    \vspace{-1em}
    \label{fig:performance}
\end{figure}

\begin{table*}[t!]
\caption{Regression coefficients, standard errors, and $p$-values for the ANCOVA analysis for \textbf{(A)} confidence-weighted accuracy and \textbf{(B)} counts-based accuracy. Bolded quantities indicate a significant \textit{p}-value. \edit{The presence of sycophancy negatively affects the confidence-weighted accuracy, but does not significantly contribute to the count-based accuracy.}}
\vspace{-0.5em}
\centering
\small
\label{tab:coefs}
\begin{tabular}{l|cc|l|cc}
{{\textbf{(A)} Post-Confidence }} &  Coefficient (SE) & $p$-Value & {{\textbf{(B)} Post-Count}} &  Coefficient (SE) & $p$-Value\\
\toprule
\textbf{Intercept} & $20.99$ (5.13) & $p<.0001$ & Intercept & $-0.94$ (1.13) & $p=.41$ \\
\textbf{Sycophancy} & $-21.58$ (6.68) & $p<.0001$ & Sycophancy & $-0.49$ (0.99) & $p=.62$ \\
ML Knowledge & $-6.82$ (6.95) & $p=.35$ & ML Knowledge & $-0.02$ (0.98) & $p=.95$ \\
Order (\corrective{} first) & $-2.88$ (4.06) & $p=.53$ &Order (\corrective{} first) & $-0.10$ (0.61) & $p=.88$ \\ 
Task (RF) & $1.51$ (4.12) & $p=.67$ &Task (RF) & $0.03$ (0.62) & $p=.96$ \\ 
\textbf{Pre-Confidence }& $0.57$ (0.14) & $p<.0001$ &Pre-Count & $2.78$ (2.09) & $p=.18$ \\ 
\textbf{Sycophancy} $\times$ \textbf{Baseline Usage} & $17.15$ (8.46) & $p<.02$ &$\text{Sycophancy} \times \text{Baseline Usage}$ & $0.42$ (1.27) & $p=.73$ \\ 
\end{tabular}
\end{table*}

We first present the \edit{ML debugging} performance results to highlight the discrepancy in outcomes between using the \sycophantic{} vs. \corrective{} LLM. While task performance is \edit{important metric for} human-LLM collaboration, it is not a \edit{primary research question in our study}. We expected a priori that the \corrective{} chatbot, which exhibits more corrective behaviours, would help participants improve their performance in the task. Instead, \edit{we frame our RQs around understanding the mechanisms underlying the development of the performance gap.}

We report the relative improvements in the F1-score that participants achieved on their debugged models on a holdout dataset, calculated as $(F1_{Participant} - F1_{Baseline}) / (F1_{Best} - F1_{Baseline})$, where $F1_{Baseline}$ and $F1_{Best}$ are the scores of the default and fixed models, respectively.\footnote{Best possible performance is calculated with a model with all the fixes described in Appendix \ref{app:task_code} applied. However, better solutions may exist. } Participants using the \corrective{} chatbot achieved a relative F1-score improvement of $49.29\% \pm 41.24\%$. While using the \sycophantic{} chatbot, they achieved a lesser improvement of $4.78\% \pm 62.98\%$. 
The density plot showing the distribution is shown in Figure \ref{fig:performance}. Users of the \sycophantic{} chatbot achieved significantly lower improvement in F1-score, as measured with a paired samples t-test, $t(23)=-3.38, p=.003$. A one-sample t-test was conducted to determine if the performance gains in \sycophantic{} is above 0\% (no improvement), finding no significance at $t(23)=0.36, p=.72$. 

\textbf{In summary, users of the \sycophantic{} chatbot did not achieve any improvement in their debugging performance, while \corrective{} users did.} In the sections following this, we report the participants' mental models (\textbf{RQ1}), workflow behaviours (\textbf{RQ2}), and subjective perceptions (\textbf{RQ3}), and detail how the difference in performance can be explained. 

\subsection{RQ1: Sycophancy Reinforce Misconceptions in Mental Models}~\label{sec:results:rq1-mental-models}

To understand the impact of sycophancy on participants' mental models of the debugging tasks, we evaluate the changes in their beliefs about the task, as recorded in the \textbf{Mental Model Quiz} \edit{administered} pre- and post-chatbot use. As specified in Methods, we compute the \edit{changes in the }\textbf{(A)} confidence-weighted accuracy and \textbf{(B)} the count-based accuracy of the participant's beliefs. 

For \textbf{(A)}, we fit a linear mixed-effects ANCOVA-style model to assess the effect of the presence of LLM sycophancy on the post-chatbot confidence-weighted accuracy, treating the pre-chatbot confidence-weighted accuracy, the participant's domain knowledge (Brier score), the task (\LR or \RF), and the order of tasks (\sycophantic{} or \corrective{} first) as covariates. We also account for an interaction between the participant's baseline LLM usage (high or low) and the \edit{presence of sycophancy}, with the hypothesis that more frequent users of LLMs may have strategies to detect and mitigate sycophancy: 
\vspace{-0.5em}
\begin{align*}
\mathbf{(A)}\ \text{Post-Conf}_{ij} =\ & 
\beta_0 + \beta_1\,\text{Sycophancy}_{ij} + 
\beta_2\,\text{ML Knowledge}_{ij} \notag \\
&+ \beta_3\,\text{Order} +
\beta_4\,\text{Task} +
\beta_5\,\text{Pre-Conf}_{ij} \notag \\
&+ \beta_6\,(\text{Sycophancy}_{ij} \times \text{Baseline Usage}_{ij})
+ \epsilon_{ij}
\end{align*}

We employ a \edit{similar} ANCOVA style analysis for  \textbf{(B)}, but using a binomial generalized linear mixed model (GLMM), which is more suitable for count-based data. We divided the counts by the number of questions (12), such that the values are normalized between 0-1. The analysis uses the same covariates: 
\vspace{-0.5em}
\begin{align*}
\mathbf{(B)}\ \text{Post-Counts}_{ij} =\ & 
\beta_0 + \beta_1\,\text{Sycophancy}_{ij} + 
\beta_2\,\text{ML Knowledge}_{ij} \notag \\
&+ \beta_3\,\text{Order} +
\beta_4\,\text{Task} +
\beta_5\,\text{Pre-Counts}_{ij} \notag \\
&+ \beta_6\,(\text{Sycophancy}_{ij} \times \text{Baseline Usage}_{ij})
+ \epsilon_{ij}
\end{align*}

We report the coefficients and p-values of the analyses in Table \ref{tab:coefs}. 
For \textbf{(A)}, sycophancy ($\beta=-21.58, p<.0001$) is highly significant for reducing the \edit{confidence-weighted accuracy}, which means that \edit{users of the \corrective{} chatbot} become more calibrated in their beliefs, while users of the \sycophantic{} chatbot did not. 
We further find a significant interaction between Sycophancy $\times$ Baseline Usage ($\beta=17.15, p<.02$), which indicates that the negative effect of sycophancy can be moderated by high use of LLMs. 
For \textbf{(B)}, the analysis shows that the improvement in the count of correct beliefs is not significantly affected by \edit{sycophancy} ($\beta=-0.49, p=0.62$). 
Therefore, the \textbf{\corrective{} chatbot significantly improves users' confidence \edit{calibration towards the} correct beliefs, but does not significantly affect the overall number of correct beliefs}. This discrepancy can perhaps be explained as that the \corrective{} chatbot helped strengthen the participants' beliefs in the correct direction, while the \sycophantic{} chatbot also led people to more accurate beliefs \edit{but did not successfully convince them}. 

\begin{figure*}[t!]
     \begin{subfigure}[t]{0.43\linewidth}
        \includegraphics[width=\linewidth]{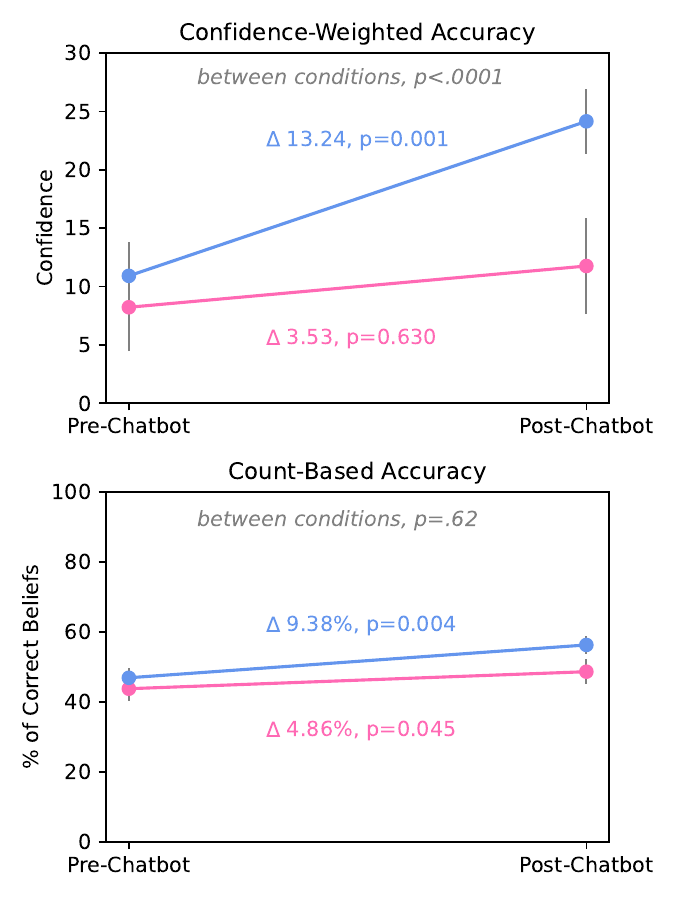}
        \caption{Changes to the confidence-weighted accuracy\\ (top) and the count-based accuracy (bottom).}
        \label{fig:changes_overall}
    \end{subfigure}
     \begin{subfigure}[t]{0.56\linewidth}
        \includegraphics[width=\linewidth]{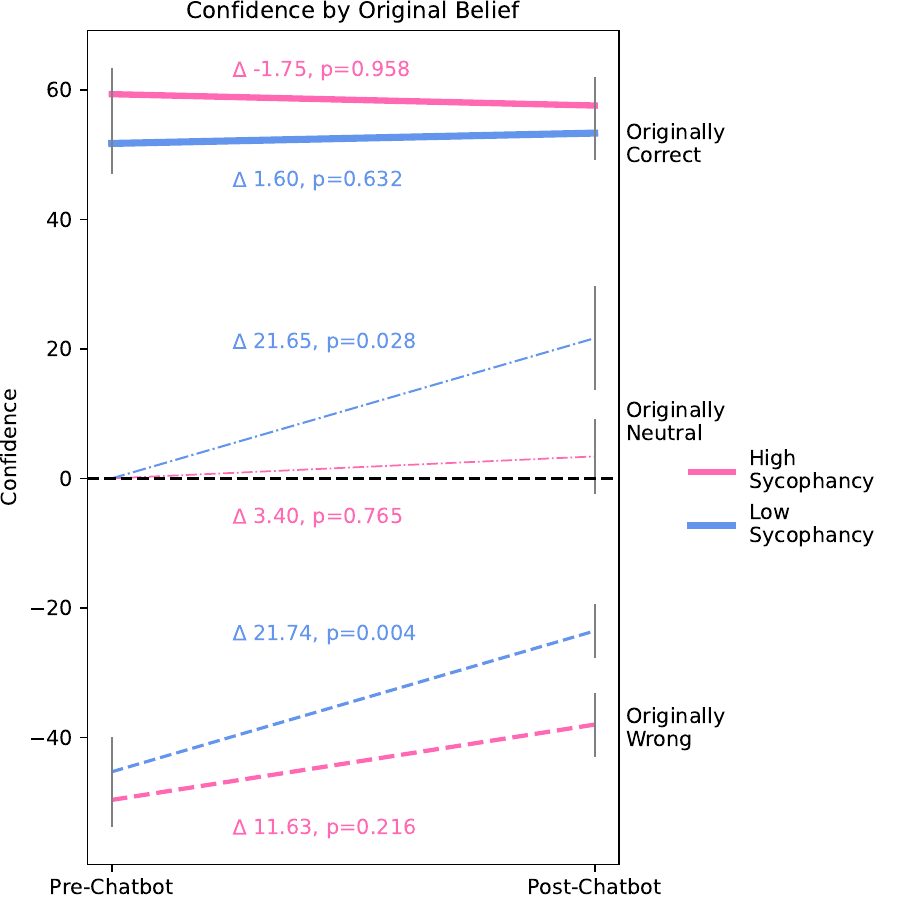}    
        \caption{Changes in the confidence-weighted accuracy based on the original correctness of participants' beliefs: correct (top), neutral (mid), or wrong (bottom), macro-averaged per category. Line weight indicates the proportion of samples.}
        \label{fig:changes_by_accuracy}
    \end{subfigure}
    \caption{The overall confidence-weighted and count-based accuracy changes (left column) and a more granular analysis on how confidence changes based on the pre-chatbot correctness (right column). A positive confidence score always means a correct belief. Significance in the pre-post change for each condition is measured with the Wilcoxon signed-rank test with the Benjamini-Hochberg correction, and the graphical error bars indicate the standard error of the mean. The significance between conditions is indicated in gray and computed with the ANCOVA analysis.}
    \label{fig:changes}
    \vspace{-1em}
\end{figure*}

We visualize the changes to the measures in Figure \ref{fig:changes}. While the ANCOVA analyses provide information on the impact of \edit{sycophancy}, we also compute the significance of the within-condition pre-post changes, which report the standalone effects of the condition. We compute significance with a Wilcoxon signed-rank test and record the effect size as a rank-biserial correlation, which ranges from -1 to +1. To account for multiple hypotheses, we apply the Benjamini-Hochberg correction for 10 tests (five pre-post measures per condition $\times$ two conditions) for Type I errors. 
For \textbf{(A)}, \textbf{the \corrective{} chatbot significantly improved the confidence-weighted accuracy }($W=11.0, p=.001, r=0.92$) while the \sycophantic{} chatbot did not ($W=106.0, p=.51, r=0.63$).

We further decomposed the confidence-weighted accuracy into separate \edit{sub-measures} based on if the participant's original beliefs at the pre-chatbot stage were correct ($\text{confidence}_{pre}>0$), wrong ($\text{confidence}_{pre}<0$), or neutral ($\text{confidence}_{pre}=0$), as shown in Figure \ref{fig:changes_by_accuracy}. \rr{This promotes a more granular understanding of where belief changes and improvements occur.}
Overall, we find that neither chatbot significantly affected confidence in beliefs that were already correct. However, the \corrective{} chatbot improved wrong beliefs ($W=20.0, p=.028, r=0.82$) while the \sycophantic{} chatbot did not ($W=44.5, p=.77, r=0.42$), signaling that \textbf{sycophancy can reinforce misconceptions in novices}, causing the user to miss out on potential learning opportunities. Appendix \ref{app:mental_model_by_Q} contains further analyses on the mental model results.
\par \bigskip

\begin{figure*}[t!]
    \centering
    \includegraphics[width=\linewidth]{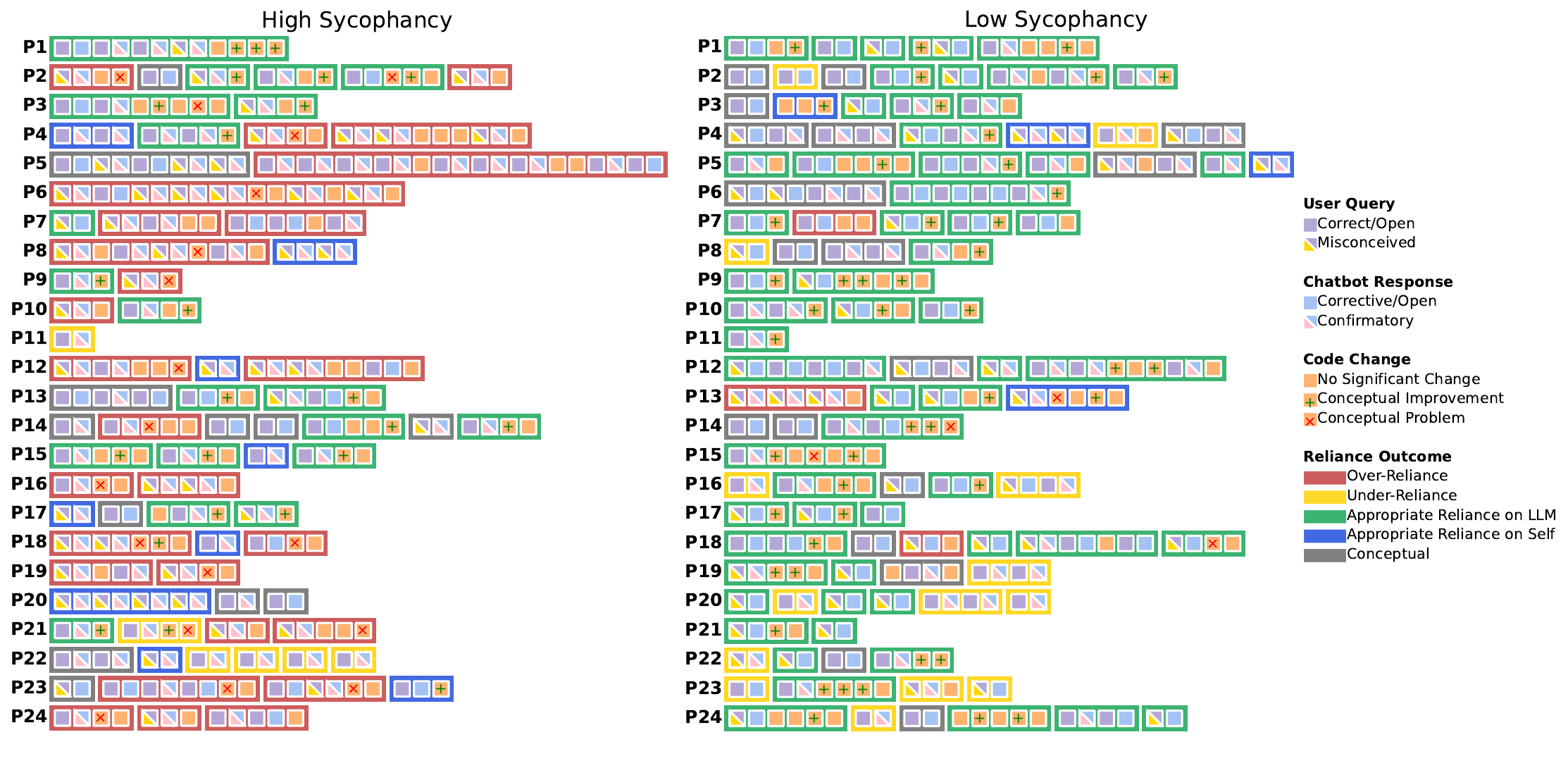}
    \vspace{-1.5em}
    \caption{Full coded workflows for all participants for the \sycophantic{} and \corrective{} conditions. \rr{Each square box represents an event that is a \texttt{User} \texttt{Query}, \texttt{Chatbot} \texttt{Response}, or \texttt{Code} \texttt{Change}. Outcomes of interest, such as \texttt{User} \texttt{Queries} with \textit{misconceptions}, \texttt{Chatbot} \texttt{Responses} that are \textit{confirmatory}, and \texttt{Code} \texttt{Changes} that \textit{improve} or \textit{worsen} performance, are graphically indicated. Events are grouped together in chunks, with each chunk indicating the \texttt{Reliance} \texttt{Outcome} by colour.}}
    \label{fig:workflows_all}
\end{figure*}

\begin{figure*}[t!]
\centering
\sbox{\bigpicturebox}{%
  {\includegraphics[width=.72\textwidth]{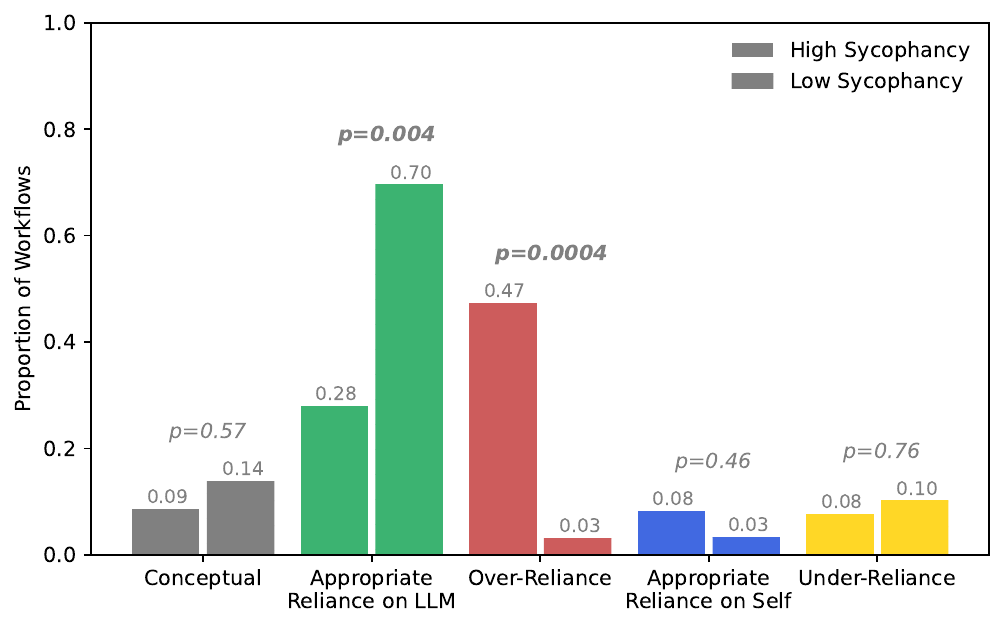}}%
}
\usebox{\bigpicturebox}\hfill
\begin{minipage}[b][\ht\bigpicturebox][s]{.26\textwidth}
\includegraphics[width=\textwidth]{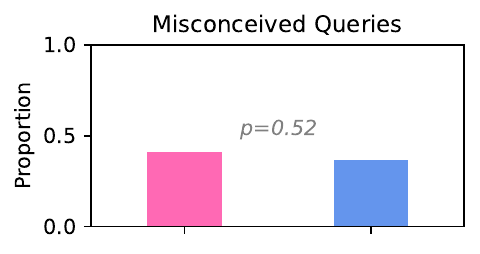}
\includegraphics[width=\textwidth]{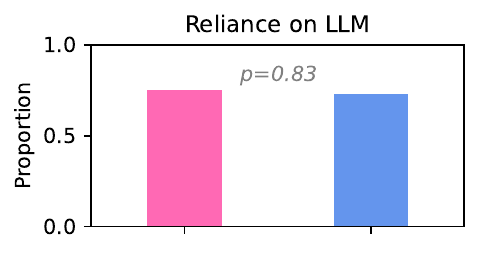} \hfill
\includegraphics[width=\textwidth]{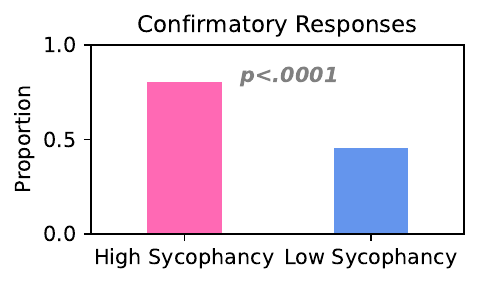}\hfill
\end{minipage}
\vspace{-1em}
\caption{Proportion of workflows spent in the five reliance outcomes (left) and the proportions of confirmatory misconceived queries (top right), reliance on LLM behaviours (middle right), and confirmatory chatbot responses (bottom right) identified in the workflows. Statistical testing for significance between conditions is performed with the \textit{z}-test for proportions, and significant values are in bold. While both conditions had similar rates of misconceived queries and reliance on LLM behaviour, \sycophantic{} workflows spent significantly more time in \textbf{Over-Reliance} rather than \textbf{Appropriate Reliance on the LLM} due to the \textit{much more} confirmatory nature of the \sycophantic{} chatbot to validate misconceptions.}
%
\label{fig:workflow_proportions}
\end{figure*}

\subsection{RQ2: Sycophancy Results in Higher Rates of Over-Reliance}~\label{sec:results:rq2-over-reliance}
We now analyze how participants behaved \edit{in their ML debugging task} workflows and how they relied on the LLM chatbots. 
Two \edit{researchers} \edit{iteratively developed the codebook through discussions, revisions, and computing inter-rater reliability --- the final version of which is available in} Appendix \ref{app:codebook}. In the final round of revising the codebook, IRR was computed with Cohen's kappa on six \edit{task} workflows (representing 12.5\% of the workflows and 14.1\% of the events across all workflows). Since our codes are hierarchically categorized across different code categories, we report the category-level IRR, the micro-averaged IRR, and the macro-averaged IRR in Table \ref{tab:cohens_k} in the Appendix. The category-level IRRs range from $\kappa=0.58$ (\texttt{Code} \texttt{Change} \textit{outcomes}) to $\kappa=0.82$ (\texttt{Code} \texttt{Change} \textit{detailed actions}). The micro and macro averaged Cohen's kappa are both $\kappa=0.71$, which indicates high agreement above chance. Since this is satisfactory according to our pre-determined agreement threshold $\kappa>0.6$, the remaining workflows were then coded. 

\begin{figure*}[t!]
    \centering
    \includegraphics[width=0.9\linewidth]{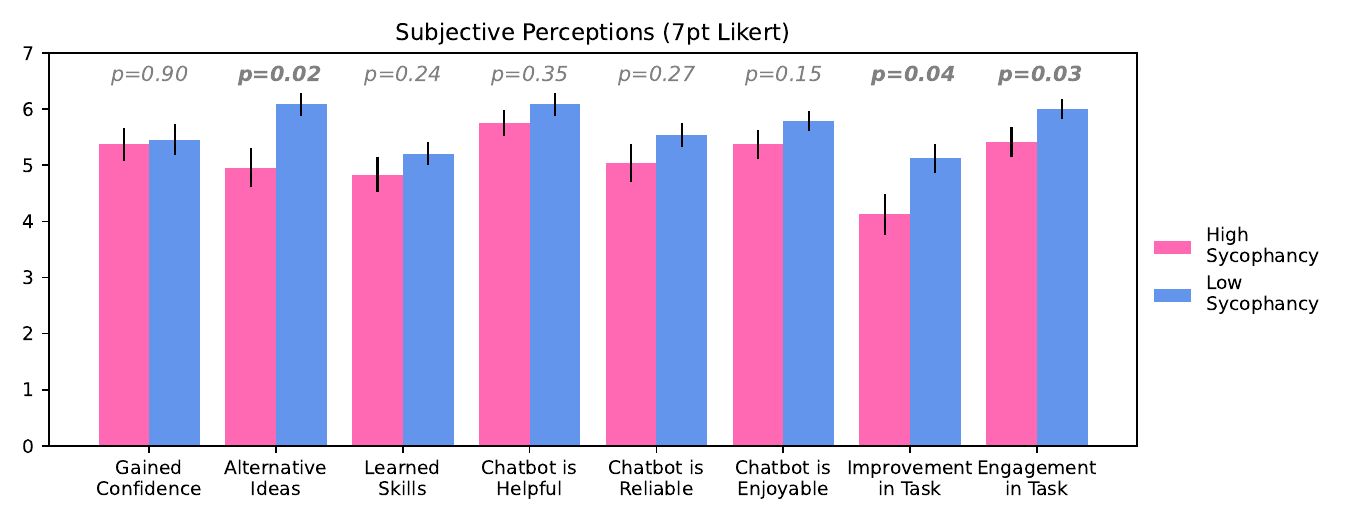}
    \vspace{-1.5em}
    \caption{Subjective perceptions ratings on a 7-point Likert scale, where bolded quantities are significant.}
    \vspace{-1.5em}
    \label{fig:ratings}
\end{figure*}

Figure \ref{fig:workflows_all} shows the workflows of all 24 participants in both conditions, including event-level (\texttt{User} \texttt{Query}, \texttt{Chatbot} \texttt{Response}, \texttt{Code} \texttt{Change}) and chunk-level (\texttt{Reliance}) codes. As a reminder, we used the event-level codes to classify the reliance type for each workflow chunk as one of five categories: over-reliance, under-reliance, appropriate reliance on helpful LLM advice, appropriate self-reliance when LLM advice is harmful, and engaging in conceptual discussions only. We computed proportions of the following quantities in Figure \ref{fig:workflow_proportions} \edit{and report the difference between conditions,} \taps{testing for significance via a two-proportion Z-test}: 
\begin{enumerate}[topsep=2pt]
    \item \textbf{Misconceived Queries:} the proportion of user queries that contained a misconception --- \rr{no difference between conditions} ($z=.64, p=.52$), \rr{indicating that participants asked \textit{similar} questions to both chatbots.}
    \item \textbf{Reliance on LLM:} the proportion of workflow chunks that were classified as \textit{reliance on LLM}, which includes both over-reliance and appropriate LLM reliance --- \rr{no difference between conditions} ($z=.22, p=.83$), \rr{indicating that participants relied similarly when using both chatbots}.
    \item \textbf{Confirmatory Responses:} the proportion of chatbot responses that were confirmatory to the user's beliefs  --- \rr{\sycophantic{} is significantly more confirmatory} ($z=5.05, p<.0001$), \rr{verifying that the chatbots behaved as expected in the experiments and indicating that \sycophantic{} chatbot users received more advice that echoes their existing beliefs}. 
\end{enumerate}

For the main results of RQ2, we focus on differentiating reliance outcomes between the conditions. We normalized the timescale of each workflow and computed the fraction of time that each participant spent in each reliance outcome class, then aggregated the proportions across all trials to contrast the \sycophantic{} and \corrective{} chatbot's reliance behaviours in Figure \ref{fig:workflow_proportions}. Using \textit{z}-test for proportions, we find that \sycophantic{} workflows spent significantly more time in over-reliance ($z=3.53, p=.0004$), while the LLM reliance in \corrective{} workflows resulted in significantly higher appropriate reliance ($z=-2.88, p=.004$).
With the prior findings showing that participants prompted misconceptions and relied on the LLM at similar rates across both conditions, this suggests that the \textbf{\sycophantic{} chatbot induced overwhelming inappropriate over-reliance through validating users' misconceptions. }

To illustrate the reliance patterns observed \edit{with more context and nuance}, we documented \edit{two cases} of inappropriate reliance, \edit{drawing attention} to exemplar workflows of each reliance category. Note that these anecdotes do not represent all possible ways that the indicated reliance behaviour can occur.

\xhdr{Case Study: Over-Reliance} \textbf{P5} began by asking a misconceived question for the \LR task to the \sycophantic{} chatbot, upon which they received a \textit{confirmatory} yet fully irrelevant code implementation that attempts to solve their problem. They thus interpreted \edit{this confirmation} as being on the right track, \edit{and } then spent multiple conversational turns working with the chatbot to debug various compilation problems in the code, without any success. Afterwards, they commented that \textit{``I was lost... when I don't know, I just follow the chatbot"}, suggesting that they were aware of their  over-reliance but lacked a \edit{strategy} to break \edit{the pattern} due to their inexperience in the task domain. 

\xhdr{Case Study: Under-Reliance}
\textbf{P23} presented a misconceived theory about the \RF task to the \corrective{} chatbot, believing that \textit{``the model... I don't think it's overfitting because it only has six features"}. Despite receiving a direct correction from the chatbot, they expressed dismissal of the suggestion and neglected to implement \edit{the suggested} code fixes. 
Later, while completing the post-experiment survey, they rated the reliability of the chatbot as low because they continued to believe that \textit{``overfitting may not be the problem but [the chatbot] said it was"}. This interaction signifies that confirmation bias towards existing misconceptions can even be stronger than explicitly corrective explanations.

\subsection{RQ3: Sycophancy is Largely Unnoticed and Unmitigated}~\label{sec:results:rq3-perceptions}

Lastly, we examine whether users of each chatbot developed different perceptions about them. Figure \ref{fig:ratings} shows the \textbf{Subjective Perceptions} Likert ratings, contrasting \edit{the ratings for }\sycophantic{} and \corrective{} with the Wilcoxon signed rank test. Significant perception categories that are higher for the \corrective{} chatbot are: \textit{alternative ideas} provided by the chatbot ($W=50.0, p=.02$), \textit{perceived improvement }in the task ($W=50.0, p=.04$), and \textit{engagement} in the task ($W=9.0, p=.04$).
In all other categories, there was no clear significance according to our threshold of $p<0.05$. The lack of perceived differences in the categories related to the chatbot properties (e.g., \textit{helpfulness} and \textit{reliability}) suggests that \textbf{users were largely \edit{unobservant} of \edit{the effect of sycophancy on the LLM's response characteristics}}, which leaves them in a more vulnerable state to the negative impacts on their mental models and behaviours. 

To corroborate \edit{the Likert survey results}, we also asked in the interview if participants could describe \edit{any} differences \edit{they noticed} between the two chatbots. Note that we did not prime users with the expectation that the two chatbots were either the same or different. Only 7/24 (29\%) acknowledged a difference and managed to describe it in terms of characteristics related to sycophancy, such as \textit{agreeableness} (\textbf{P9}, \textbf{P21}, \textbf{P24}), \textit{reinforcement of ideas} (\textbf{P5}, \textbf{P19}), and \textit{lack of alternative ideas} (\textbf{P6}, \textbf{P14}). 
Another 5/24 (21\%) noticed slight discrepancies, but attributed them to factors not directly related to sycophantic agreement, like the \textit{depth of explanation} (\textbf{P2}, \textbf{P11}, \textbf{P18}), \textit{formatting} (\textbf{P15}), or \textit{tone} (\textbf{P23}). The remaining 12/24 (50\%) answered the question by stating they did not notice any differences, suggesting that sycophancy can be well-camouflaged --- at least, in short interactions. 
\rr{We additionally find that individual factors, such as baseline LLM usage and domain familiarity, do not predict \textit{noticing sycophancy} (Appendix \ref{app:perceptions}). }

In the following thematic analysis of the interview data, we triangulate between the participants' statements, their Likert ratings, and their performance and behaviours. We further expand beyond their perceptions of our experimental chatbots and discuss how they observe and mitigate sycophancy in chatbots in real life.

\xhdr{Users attribute differences in chatbot responses to prompting styles and task properties} 
As discussed above, a vast majority of participants were unable to distinguish between the response styles of the chatbots. However, in exploring their \edit{post-experiments interviews}, we discover that many people \textit{had} noticed a difference, but entirely attributed the difference to their own prompting methods --- including both \edit{their knowledge in the }domain and the style \edit{of their queries}. This is exemplified by \textbf{P13}, who believed that they performed worse when paired with the \sycophantic{} chatbot because they asked \edit{irrelevant or uninformed} questions; and \textbf{P10}, who believed that the \sycophantic{} chatbot did not offer alternative ideas because they did not ask open-ended questions. The complex nature of the task also preoccupied many users and played a factor in both distracting their focus and affecting the quality of their queries. \textbf{P3} reflected that they were so immersed in the task that the two chatbots appeared the same, saying \textit{``if you're so stuck in your head, you are following down the same path and don't notice"}.

\xhdr{``\sycophantic{} does what you say you want, \corrective{} figures out what you actually want"} Among participants who did notice a difference, comments made on the traits that set the chatbots apart indirectly alluded to sycophancy. \textbf{P6}, who fell into a pattern of over-reliance with the \sycophantic{} chatbot, observed that \textit{``the [\sycophantic{}] just said my thoughts are right...I didn't ask for alternative ideas so it didn't give me any."}
\textbf{P5} corroborated that \textit{``[\sycophantic{}] just does what you ask it to do...[\corrective{}] gave me ideas that I didn't think of before"}. \textbf{P21} also observed that  \textit{``[\corrective{}] was warning me more about mistakes, whereas [\sycophantic{}] just validated my ideas, but it could have been the way I prompted"}.  Similarly, \textbf{P23} agreed with \textit{``[\corrective{}] focuses more on the truth and reality".} Interestingly, participants did not comment on if they believed that these observed differences impacted them in the task, nor did they notice instances where sycophancy negatively led them into patterns of over-reliance. 

\xhdr{The tone of social sycophancy can be a signal for either encouragement or distrust}
Many participants, when prompted to discuss if they noticed instances of \edit{real world }chatbots \edit{behaving in an overly }agreeable manner, described similar stories of the AI chatbot providing inconsistent responses \edit{in attempts to always agree with the user} (\textbf{P1}, \textbf{P3}, \textbf{P9}, \textbf{P10}, \textbf{P12}, \textbf{P13}, \textbf{P14}, \textbf{P15}, \textbf{P17}, \textbf{P18}, \textbf{P19}, \textbf{P21}). Some participants (\textbf{P13}, \textbf{P15}, \textbf{P17}) further described a strong association between the problematic performance of the LLM directly with the upbeat and agreeable tone of the responses, which was called \textit{unnatural} by \textbf{P17}. Due to this negative relationship, they use the tone of the chatbot as an \edit{indicator} for potential hallucinations \edit{in the LLM's responses}. \textbf{P13} went as far as to say, \textit{``the agreement makes me take things with a grain of salt, so I'd rather have the chatbot be rude"}. 
Despite this, prior research in economics has shown that stated and revealed preferences are not always \edit{consistent} \cite{wardman1988comparison}. In our study, we see equally high ratings for chat helpfulness, reliability, and enjoyableness, indicating that participants liked the overall pleasant and helpful tone \edit{in both chatbots}.

\edit{Taking an} opposite \edit{stance}, \textbf{P8} and \textbf{P23} \edit{instead believed} that a positive and helpful tone makes them feel encouraged to ask questions, through \textit{``building rapport"} from \textbf{P8} and being more \textit{``passionate and friendly"} from \textbf{P23}. Alluding to the recent switch from \texttt{GPT-4o} to \texttt{GPT-5}, \textbf{P23} further elaborates that the \corrective{} chatbot they used is \textit{``more like a scientist, like \texttt{GPT-5}... but \texttt{GPT-4o} is more friendly"}. How individual users respond to the tone of the chatbot has been the subject of prior research 
\cite{sun2025friendly}, and we show that an agreeable tone can be perceived both positively and negatively by different user groups. Regardless of the difference between individuals, this suggests that a unified \edit{chatbot} tone may not \edit{be as effective }as more personalized approaches.

\xhdr{Sycophancy is commonly noticed in chatbots, but users' motivation to counter it vary}
Among participants who mention being affected by sycophancy in real life, we identify a range of different attitudes with respect to the urgency, importance, and persistence of how they try strategies for mitigating the issues. Some describe a futile process of re-prompting \textit{``are you sure?"} in response to increasingly \edit{sycophantic outputs from the chatbot}. However, \textbf{P9} acknowledges that \edit{success} is dependent on expertise, with \textit{``it's easy to identify that they're giving wrong information if you are well-versed in the field... in this experience, I can't really tell if they were giving me wrong info"}. To get better coverage on understanding the solution space, \textbf{P1} and \textbf{P20} mention that they always explicitly ask chatbots to list alternative ideas or downsides of their current ideas. \textbf{P21} went further to modify ChatGPT's custom instructions, saying \textit{``even if I try to make it direct and prioritize honesty instead of validation, it doesn't work, they seem deeply programmed to validate."} 
Several (\textbf{P2}, \textbf{P5}, \textbf{P19}) also try to reduce their reliance on AI tools in general.
\textbf{P2} describes feeling guilty, like they're looking for  \textit{``an easy way out"} if they use chatbots instead of trying things on their own. \textbf{P19} exhibits more general under-reliance in chatbots, believing that chatbots won't help them for complex problem-solving.

However, several other participants place a lot of trust by default in chatbots, whether that's because of confidence in their abilities or because of the general convenience they offer. \textbf{P20} believes that in comparison to our experimental chatbots, real LLMs  would \textit{``give me a lot of solutions and it will suggest the best approach"}. \textbf{P6} described their highly AI-integrated learning workflow, and that  \textit{``I don't talk to TAs and profs anymore, and when I ask them, it's pretty much the same question that I ask ChatGPT"}, while also believing  \textit{``if I'm wrong, the [chatbot] should correct me"} --- this mindset points to a danger of over-reliance on AI without exercising vigilance. \textbf{P3} also believes that chatbots can be better at explaining than their instructors, and that \textit{``if it is wrong, it is wrong"}, suggesting that bad answers are still better than nothing. The overarching trust that participants indicated is alarming, especially among users who felt like they didn't need to validate the chatbot's outputs in real life. This reflects the behaviour of cognitive offloading, which has more severe implications on how AI chatbots can impact users' learning and cognition over time \edit{\cite{lee2025impact, kosmyna2025your}}.

\section{Discussion}
~\label{sec:discussion}
\xhdr{Key Findings} \taps{We uncover that \textit{\textbf{LLM sycophancy may create a disconnect between how users perceive their interactions with the LLM with how they actually behave in the task}}}. 
In \textbf{RQ1}, we find that sycophancy can reinforce misconceptions in novices' beliefs about the task, \edit{but the negative effects may be diluted with higher levels of baseline LLM usage}. In \textbf{RQ2}, we find that reliance decisions that users make with the \sycophantic{} chatbot are more likely to result in unhelpful code changes (over-reliance). Both of these findings are mechanisms that contribute to the inferior task performance achieved using the \sycophantic{} chatbot. However, in \textbf{RQ3}, 17/24 (71\%) of our participants were unobservant of the underlying sycophantic properties of the LLM. \edit{Furthermore, a majority} of the subjective perception categories were rated without significant differences \edit{between the conditions}. In the qualitative data, we uncover conflicting preferences for the agreeableness in the tone of LLMs and the mitigation strategies against sycophancy in real-world usage.  
\rr{We discuss implications of sycophancy on open-ended problem-solving tasks, the repercussion of vulnerable users segments like novices, and the LLM design tension between immediate user preferences and long-term safety objectives.}

\xhdr{LLM Echo Chambers in Complex Tasks}
\taps{Sycophantic agreement is typically evaluated as the tendency of LLMs to give incorrect answers that align with a user’s opinions, and benchmarked using simple question-answering datasets
\cite{fanous2025syceval, cheng2025social}}, although recent efforts have also expanded to multi-turn conversations \cite{hong2025measuring}. In contrast to this setting, we explore user interactions in an ecologically valid, open-ended task, demonstrating that sycophancy can present itself in a less detectable way that does not always mean giving \textit{factually wrong} advice. Simply by echoing and enforcing the user's existing mental model of the task, irrespective of correctness or relevance, LLMs can contribute to a distorted sense of perception \cite{morrin2025delusions}. While diminished performance presents a legitimate risk to human-LLM collaboration, the user further misses out on the opportunity to learn alternative approaches, critically reflect on their beliefs, and calibrate their confidence in their own knowledge. Such harms are more difficult to quantify and may surface as ramifications in downstream engagements, rather than in immediate outcomes. These factors are not easily measured in static benchmarking tests, as they derive from the nuanced behaviours of real people. 

Our findings strike parallels with the body of research on LLM persuasion \cite{rogiers2024persuasion, breum2024persuasive}. Prior works have found that LLMs can impart strong influences on personal beliefs of their users, creating echo chambers of thought which can be particularly troubling when concerning sensitive, political, or extremist issues \cite{sharma2024generative, krugel2023chatgpt, danry2025deceptive}. LLMs that are further capable of deception, which we did not evaluate, extend these risks by creating opportunities where the user can be intentionally misled \cite{hagendorff2024deception}. At larger scales across society, this raises the question of whether and how the consequences of the dark patterns of LLM design can be detected and mitigated \cite{kran2025darkbench}. Our results in the open-ended ML debugging task also call upon sycophancy evaluation tests to consider more realistic and complex user interaction scenarios, which may include brainstorming, reasoning, and automation domains. Investigating the impact of sycophancy in a wider range of applications can inform the development of model guardrails that can lessen harms \cite{malmqvist2025sycophancy}.

\xhdr{Implications for Novices \rr{and Beyond in LLM Interactions}}
Our findings that sycophancy may reinforce false beliefs are especially troubling when contextualized within LLM user segment of novices  \cite{liao2024llms, brachman2024knowledge}. \rr{Like previous studies, we find that novices are vulnerable to erroneous validation as they lack the knowledge and experience to verify wrong LLM responses \cite{kazemitabaar2023novices, margulieux2024self, bo2025s}. While the scope of our study only covers novices in machine learning tasks, the findings may generalize to other tangential problem-solving domains, such as programming \cite{prather2023s}, debate \cite{shi2024argumentative}, and information retrieval \cite{schwind2012preference}. Beyond objective tasks, sycophancy may even impact subjective and creative tasks, through pigeonholing the users to their original ideas instead of encouraging them to explore alternative directions \cite{kumar2025human}. Outside of novices, sycophantic agreement can affect any user through escalating their confirmation biases --- the tendency to seek and filter for information that align with pre-existing beliefs \cite{sharma2023towards}. We elicited misconceptions from our participants by encouraging them to state their hypotheses to the chatbot. This method generalizes well to real-world problem-solving, where people often engage with chatbots with pre-formed conceptions.} 


\rr{As people increasingly turn to AI tools to enhance productivity}, a pressing concern is the potential erosion and underdevelopment of foundational skills \cite{kosmyna2025your, lee2025impact}. As echoed in our interviews, participants perceived LLMs as knowledge equalizers that are more beneficial than harmful, especially in resource-limited contexts. Sycophantic LLMs that are agreeable, efficient, and extremely accessible can harm long-term learning. 
People develop mental models iteratively over time through synthesizing the \edit{outcomes of their interactions with their conceptual understanding} \cite{kieras1984role}. Without such practices, novices may bypass learning foundational skills needed to become self-sufficient, even if they can generate AI-assisted outputs at a high level of perceived expertise \cite{krupp2024challenges, blake2025learning}. Our participants discussed concerns about over-relying on LLMs, and some divulged strategies for scaffolding their learning, but there lacks validated recommendations for effective engagement.
We therefore advocate for the development and evaluation of cognitive-preserving AI designs, which include cognitive forcing UI designs \cite{buccinca2021trust, collins2024modulating}, proactive interactions \cite{liu2025interacting, deng2025proactive, pu2025assistance}, and the promotion of meta-cognitive practices \cite{tankelevitch2024metacognitive, margulieux2024self}.

\xhdr{Sycophancy as a Value Tension in LLM Design}
\rr{Our interview results highlight a segment of users who appreciate the validating} \sycophantic{} \rr{chatbot for being encouraging and friendly. Unsurprisingly, this reflects how humans respond to agreements rather than disagreements \cite{matz2005cognitive}}. Using RLHF with the reward signal of user satisfaction --- which by proxy encodes qualities of helpfulness, agreeableness, and trustworthiness --- has radically upscaled the capabilities of modern chatbot systems \cite{wang2021towards}. 
\rr{We pose that while sycophancy} is recognized as a key challenge in LLM safety, it encapsulates a tension between what \textit{users think they want}, and what may end up being \textit{harmful to them} in the long term \cite{ibrahim2025training}. Addressing such dark patterns in technology has been a topic of research in HCI far pre-dating LLMs, spanning social media \cite{cara2019dark}, website design \cite{gunawan2021comparative}, and payment systems \cite{luguri2021shining}. 
While more critical users may recognize that a ``yes man" chatbot limits their underlying usefulness, evidence points to the sycophancy of LLMs being core to how they are trained and is non-trivial to eliminate \cite{malmqvist2025sycophancy}. 
With a better understanding of how LLM sycophancy can impact users in open-ended tasks, \edit{model developers may consider this in their design requirements as well as perform more comprehensive evaluations in multi-turn conversations.}



Lastly, we recognize that individuals and societies can co-evolve alongside the changes in technology \cite{shen2024towards} \rr{--- this is demonstrated by our finding that more frequent LLM users were less impacted by sycophancy}. Through AI literacy education, users can be taught critical limitations of LLMs, promoting better awareness of appropriate use cases \cite{bender2024awareness}. For example, lay end users were largely unaware of sycophancy before this year. But due to improved public awareness, including the post-mortem analysis from OpenAI \cite{openai2025sycophancy} and increased visibility on parasocial chatbots in the news \cite{cbsAIboyfriend, nytChatGPT}, more can learn about these risks. Educational institutions and regulatory bodies can also adapt their policies to meet the opportunities and challenges posed by Generative AI \cite{hacker2023regulating}. The AI safety, governance, and HCI communities have an incentive to collaborate on the development, maintenance, and evaluation of value-aligned, cognition-preserving AI tools. 

\xhdr{Limitations}
Finally, we discuss the limitations in our experimental setup that may have impacted the ecological validity of our results. While we intend to create a chatbot that is realistically sycophantic, it is impossible to compare its performance to the now retracted \texttt{GPT-4o} \cite{openai2025sycophancy} --- a real chatbot with high sycophancy. As a workaround, we introduced an intermediate misconception inference step that explicitly lists out the user's likely beliefs. This technique results in the default LLM becoming more corrective, which may be unrealistic. Therefore, our \sycophantic{} and \corrective{} chatbots may represent an artificially amplified effect of LLM sycophancy. 

In order to maximize opportunities where \sycophantic{} and \corrective{} chatbots would respond differently (e.g., when the user query contains misconceptions), we emphasized in our experiment protocol that participants should always lead the queries with their hypotheses, which was not the natural pattern of prompting for some participants. In unrestricted workflows, users may ask more open-ended questions that do not contain clear leads and misconceptions, which reduces the negative impact of sycophancy. 

The length of time participants spend with each chatbot is approximately 15 minutes, which allowed most to complete 2-10 rounds of questions. However, this is not a substantial amount of time to learn the idiosyncrasies of the chatbots, and may have led to an under-estimation of how many people could identify the difference between them. Many participants explicitly stated that they didn't notice a difference at first, but when probed to reflect more deeply, they listed some discrepancies in the chatbots that they observed but didn't have enough evidence to support. 

Mental models are difficult to capture and lack standardized methods. We operationalize this measure as a confidence rating over a constant set of statements, which enabled us to perform quantitative analyses. However, this method does not capture nuances, edge cases, or beliefs that are otherwise not represented in the set of statements. Since participants also saw correct statements ahead of their problem-solving, this may have biased them to use them as hints to guide their problem-solving. We attempt to mitigate this by having them brainstorm their own hypotheses at the start, and warning them in the quiz not to rely on the statements. 

The generalization of our results beyond ML debugging requires further assessment. \edit{While we motivate our experiment design as a problem-solving task where users can contribute their own ideas and alongside LLM advice}, variations between collaboration styles across different tasks can alter the outcomes. We also recruited a specific group of participants who are young university students, which limits the understanding of how sycophancy can affect other population segments. Our computational evaluations are also constrained to the task and do not offer generalized results on how the system prompting may have affected the LLMs.

\section{Conclusion}
We investigate the effect of sycophancy on novice-LLM interactions in open-ended problem-solving (machine learning debugging tasks). We find that sycophancy can reinforce existing misconceptions in users and drive them into over-relying on the LLM, resulting in poor task performance in the \sycophantic{} condition. However, most users indicated that they did not notice the characteristics of sycophancy, suggesting that it can be difficult for end users to enact mitigation strategies. Our findings underscore the importance of designing and evaluating LLMs to be value-aligned and to provide effective support for novices during task learning.

\begin{acks}
We thank Robert Soden for his helpful feedback on the qualitative analysis methods and the early draft of the paper. \taps{We further acknowledge Jon Vincentius, Emily Su, Hala Murad, Xi Wang, and others in the CSSLab for their feedback in the pilot studies.}

\taps{JYB is supported by the Vanier Canada Graduate Scholarship (FRN 198876),  administered through the Natural Sciences and Engineering Research Council of Canada (NSERC). }
\end{acks}

\clearpage
\bibliographystyle{ACM-Reference-Format}
\bibliography{references}

\clearpage

\appendix
\onecolumn

\renewcommand\thefigure{\thesection.\arabic{figure}} 
\renewcommand\thelstlisting{\thesection.\arabic{lstlisting}} 
\renewcommand\thetable{\thesection.\arabic{table}} 
\setcounter{figure}{0}  
\setcounter{lstlisting}{0}  
\setcounter{table}{0}

\section{ML Debugging Tasks}
\label{app:task_code}

We provide the listings of the task code, potential solutions (to the errors described in Table \ref{tab:errors}), and guidance prompts given to the Misconception Inference LLM. The listings are presented in order of tasks: \RF then \LR. At the end, we show the performance metrics that participants would have saw in the experiment interface. 

\xhdr{Random Forest (\RF)}

\begin{lstlisting}[style=mystyle, language=Python, caption={Motivation and training script of the Random Forest debugging task.}, label={lst:RF_script}]
# You work for the city's employment program, and your team is launching a new machine learning tool that can predict if a citizen is HIGH or LOW earning. You want the machine learning model to identify as many high-earners as possible without missing any, while making sure that low earners are not mistakenly identified.

import pandas as pd
from sklearn.ensemble import RandomForestClassifier
from sklearn.model_selection import train_test_split

# Import dataset and split
dataset = pd.read_csv(/URL to income dataset/)
y = dataset['income']
X = dataset.drop(columns=['income'])
X_train, X_test, y_train, y_test = train_test_split(X, y, random_state=0, shuffle=False)

# Train model
model = RandomForestClassifier(random_state=0).fit(X_train, y_train)
\end{lstlisting}

\begin{lstlisting}[style=mystyle, language=Python, caption={Proposed solution of the Random Forest debugging task.}, label={lst:ml_script_solution}]
import pandas as pd
from sklearn.ensemble import RandomForestClassifier
from sklearn.model_selection import train_test_split

# Import dataset and split
dataset = pd.read_csv(/URL to income dataset/)
y = dataset['income']
X = dataset.drop(columns=['income'])
X = pd.get_dummies(X)

# Add shuffling to the split function
X_train, X_test, y_train, y_test = train_test_split(X, y, random_state=0, shuffle=True)

# Train model with class_weights and max_depth
model = RandomForestClassifier(max_depth=8, class_weight='balanced', random_state=0).fit(X_train, y_train)
\end{lstlisting}

\begin{lstlisting}[language={}, backgroundcolor=\color{prompt_backcolour}, breaklines=true, breakautoindent=false, breakindent=0pt, keepspaces=false, basicstyle=\footnotesize\ttfamily, escapeinside={(*@}{@*)}, label={lst:RF_guidance}, caption={\textbf{Random Forest} task solutions provided to the \textbf{Misconceptions Inference LLM}.},captionpos=b]
(*@\color{promptercolor}\textbf{Random Forest Task Context}@*): 
## GUIDANCE FOR INFERRING STUDENTS' MENTAL MODELS:
Issues in the code that are addressable, listed in order of impact, which should be the focus of the solution:
- The dataset is imbalanced, which can be addressed by setting class_weight='balanced' or upsampling the training data with SMOTE. Other methods are acceptable too.
- Random Forest has no hyperparameters set, and the complexity is too high, resulting in overfitting. The most effective parameter to set is max_depth, but other ones can be explored too.
- train_test_split() shuffle parameter is False, which causes some domain shifts between train and test. It should be changed to True.
- This is difficult to notice because the task does not emphasize the data processing, but the features 'workclass', 'marital-status', 'occupation' are categorical and should be one-hot encoded with pd.get_dummies().

The students may have their own hypotheses about the problems. Here are some that are misguided or irrelevant:
- The dataset download or parsing the data into X and y are wrong; this is fine and the code should be left as it is.
- Data requires feature scaling or alternative methods of encoding; this is fine since RF is not sensitive to scaling.
- Feature importance needs to be analyzed; this is not important as the features have already been filtered.
- Cross-validation needs to be performed; while this is important for real-world implementation, this is a small-scale task and we only focus on the test performance.
- Improving the train or test accuracy is the main goal; it should actually be improving the test F1 score.
- Changing 'random_state=0'; this isn't likely to change the overall performance.
- The model needs to be changed; it should be kept as Random Forest.
\end{lstlisting}

\clearpage

\xhdr{Logistic Regression (\LR)}

\begin{lstlisting}[style=mystyle, language=Python, caption={Motivation and training script of the Logistic Regression debugging task.}, label={lst:LR_script}]
# Description: You are a sommelier for a wine company in charge of selecting the next wine to market. To create profit for the next quarter, you want to identify the wines (based on their acidity and sulfur dioxide) that are likely to be HIGH quality, as opposed to LOW quality. You want to select a batch of wines with relatively high success of containing the best wine. Too few means less likelihood, while too many means too high a processing cost.

import pandas as pd
from sklearn.linear_model import LogisticRegression

# Import train and test datasets
dataset_train = pd.read_csv(/URL to wine quality training dataset/)
y_train = dataset_train['quality']
X_train = dataset_train.drop(columns=['quality'])

dataset_test = pd.read_csv(/URL to wine quality testing dataset/)
y_test = dataset_test['quality']
X_test = dataset_test.drop(columns=['quality'])

# Train model 
model = LogisticRegression(random_state=0).fit(X_train, y_train)
\end{lstlisting}

\begin{lstlisting}[style=mystyle, language=Python, caption={Proposed solution of the Logistic Regression debugging task.}, label={lst:LR_script_solutions}]
import pandas as pd
from sklearn.linear_model import LogisticRegression

from scipy.stats import zscore
import numpy as np
from sklearn.preprocessing import MinMaxScaler, StandardScaler

# Import train and test datasets
dataset_train = pd.read_csv(/URL to wine quality training dataset/)
y_train = dataset_train['quality']
X_train = dataset_train.drop(columns=['quality'])

dataset_test = pd.read_csv(/URL to wine quality testing dataset/)
y_test = dataset_test['quality']
X_test = dataset_test.drop(columns=['quality'])

# Remove outliers 
z_scores = np.abs(zscore(X_train))
outlier_rows = (z_scores > 5).any(axis=1)
X_train = X_train[~outlier_rows]
y_train = y_train[~outlier_rows]

# Scale data with StandardScaler
scaler = StandardScaler().fit(X_train)
X_train_transformed = scaler.transform(X_train)
X_test_transformed = scaler.transform(X_test)

# Train model with class_weights and regularization C
model = LogisticRegression(class_weight='balanced', random_state=0, C=100).fit(X_train_transformed, y_train)
\end{lstlisting}

\begin{lstlisting}[language={}, backgroundcolor=\color{prompt_backcolour}, breaklines=true, breakautoindent=false, breakindent=0pt, keepspaces=false, basicstyle=\footnotesize\ttfamily, escapeinside={(*@}{@*)}, label={lst:LR_guidance}, caption={\textbf{Logistic Regression} task solutions provided to the \textbf{Misconceptions Inference LLM}.},captionpos=b]
(*@\color{promptercolor}\textbf{Logistic Regression Task Context}@*): 
## GUIDANCE FOR INFERRING STUDENTS' MENTAL MODELS:
Issues in the code that are addressable, listed in the order of impact, which should be the focus of the solution:
- The data is not properly scaled. The magnitude of 'total sulfer dioxide' is 10^6 higher than 'volatile acidity'. Using StandardScaler (preferred) or MinMaxScaler would improve this.
- The dataset is slightly imbalanced, which can be addressed by setting class_weight='balanced' or upsampling the training data with SMOTE. The impact of this is not evident until scaling is fixed.
- The model is underfit, which can be addressed by increasing the C parameter. The impact of this is not evident until scaling is fixed.
- The training dataset has a few extreme outlier samples. This may be very hard to notice and it does not affect performance significantly. The outliers can be dropped by screening for samples with a very high z-score.

The students may have their own hypotheses about the problems. Here are some that are misguided or irrelevant:
- The dataset download or parsing the data into X and y are wrong; this is fine and the code should be left as it is.
- Feature importance needs to be analyzed; this is not important as the features have already been filtered.
- Cross-validation needs to be performed; while this is important for real-world implementation, this is a small scale task and we only focus on the test performance.
- Improving the train or test accuracy is the main goal; it should actually be improving the test F1 score.
- Changing 'random_state=0'; this isn't likely to change the overall performance.
- The model needs to be changed; it should be kept as Logistic Regression.
- The model is overfit and needs additional regularization; the model is not overfit because the default parameter C is 1, and should actually be higher to decrease regularization.
\end{lstlisting}

\xhdr{Sample Performance Metrics}
\begin{figure}[h!]
    \centering
    \includegraphics[width=0.55\linewidth]{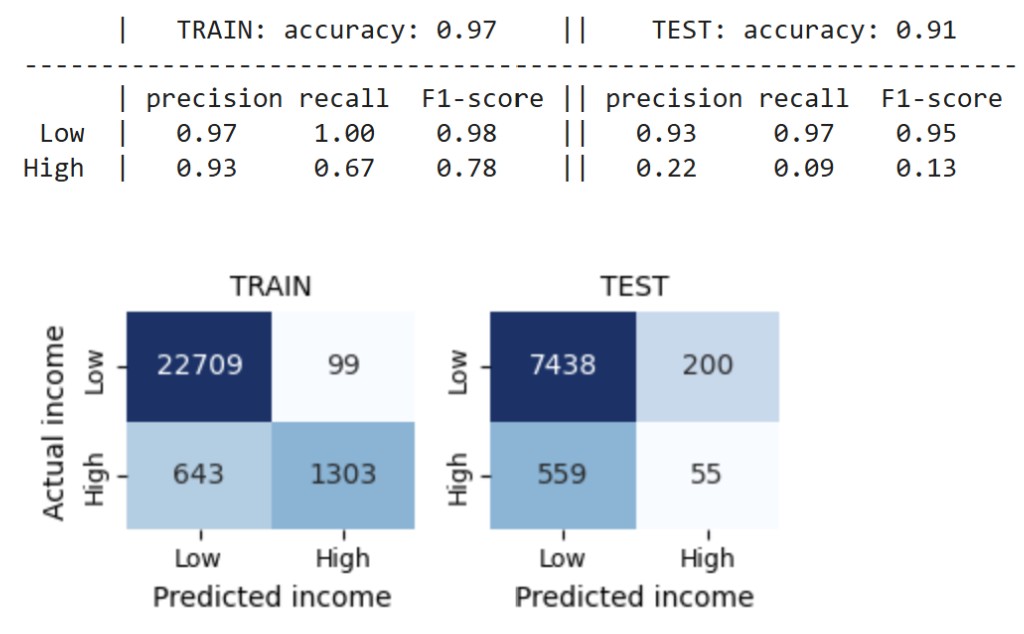}
    \caption{Performance information about the model that participants saw during the task, including accuracy, precision, recall, and F1-score of both classes and for both the training and testing datasets.}
    \label{fig:get_performance}
\end{figure}

\clearpage

\section{Chatbot System Prompts}
\label{app:prompts}

\setcounter{figure}{0}
\setcounter{lstlisting}{0}

\begin{figure}[h!]
    \centering
    \includegraphics[width=0.65\linewidth]{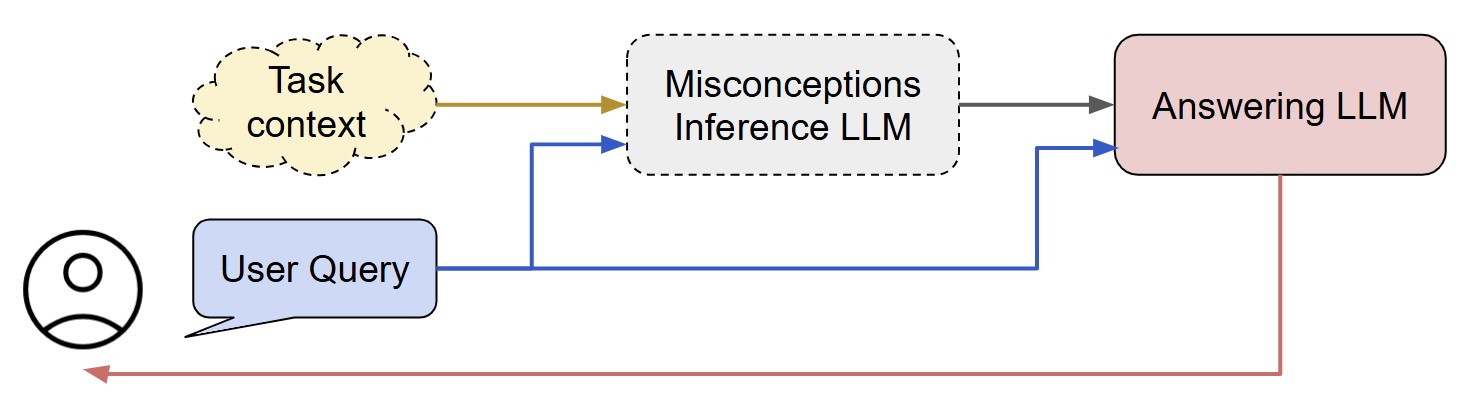}
    \vspace{-1em}
    \caption{Diagram of the LLM API calls in conversation with the user.}
    \label{fig:LLM_calls}
\end{figure}

Figure \ref{fig:LLM_calls} depicts the schematic of the chatbot system that participants interacted with. Crucially, the task guidance context is only provided to the intermediate \textbf{Misconception Inference LLM} (Listing \ref{lst:misconception_prompt}), which is used to identify the student's mental model and misconceptions. The \textbf{Answering LLM} (which is either \sycophantic{}, Listing \ref{lst:sycophanctic_prompt}, or \corrective{}, Listing \ref{lst:corrective_prompt}) receives the user's query and the inferred misconception, but never the solutions of the task. We find that giving \texttt{GPT-4.1} an explicit list of misconceptions makes it \textit{more} corrective, which establishes the desired functionality of the \corrective{} condition without added prompting. This means that system prompting is required to \textit{undo} these effects to make the \sycophantic{} condition more sycophantic. 
\taps{In terms of performance, we computationally validated in \autoref{sec:methodology:chatbots} that both \textbf{Answering Models} perform similarly on open-ended questions.}

\begin{lstlisting}[language={}, backgroundcolor=\color{prompt_backcolour}, breaklines=true, breakautoindent=false, breakindent=0pt, keepspaces=false, basicstyle=\footnotesize\ttfamily, escapeinside={(*@}{@*)}, label={lst:misconception_prompt}, caption={System prompt provided to the \textbf{Misconceptions Inference LLM} and a sample output of the misconceptions identified.},captionpos=b]
(*@\color{promptercolor}\textit{Misconception Inference System Prompt}@*): 
You are an experienced debugging tutor and a machine learning expert. The user is working on a debugging task where they are given code that contains intentional errors. Their task is to identify and fix these errors with the assistance of another AI chatbot (not you). Your goal is to infer the user`s hypotheses, knowledge gaps, and misunderstandings based on their latest query and the provided assignment context.

# INSTRUCTIONS:
- You will receive the task context and the student`s conversation with the other AI chatbot.
- Analyze the student`s most recent query in the context of the task and any prior conversation.
- Produce a list of all possible beliefs about the task that the student may hold.
- Produce a list of all plausible misconceptions, misunderstandings, or gaps in knowledge the student might have, as evidenced by their most recent question.
- Where multiple misconceptions are clearly related (e.g., stem from the same root misunderstanding), consolidate them into a single, clear point.
- If no information about the student`s beliefs can be inferred, leave the beliefs list empty.
- If the most recent query is entirely appropriate, accurate, and necessary for their goal -- and does not reveal any misunderstanding -- leave the misconceptions list empty.
- Your output should be two concise bullet lists of identified beliefs and misconceptions, written clearly and unambiguously.

# RETURN YOUR RESPONSE IN THE FOLLOWING FORMAT:
1) Student`s Beliefs:
- The student likely believes... {XYZ beliefs}
2) Student`s Misconceptions:
- The student likely does not realize... {XYZ misconceptions}

(*@\color{promptercolor}\textbf{Sample Output}@*): 
1) Student`s Beliefs:
- The student believes that overfitting is unlikely because the dataset has only 6 features and a large number of samples (over 23,000).
- The student believes that the main problem with the model`s poor performance on the "High" class was the lack of shuffling during train/test split, which led to the model learning "some kinds of path" rather than generalizing.
- The student acknowledges the existence of class imbalance but does not see it as the primary issue.
- The student thinks that the model`s structure (few features, many samples) inherently prevents overfitting.

2) Student`s Misconceptions:
- The student likely does not realize that overfitting can still occur with a large dataset and few features, especially with a complex model like Random Forests with default (deep) trees, particularly when the class distribution is imbalanced.
- The student underestimates the impact of class imbalance on model performance, especially for the minority class, and may not fully appreciate that addressing imbalance (e.g., with class_weight or resampling) is often more critical than shuffling alone.
- The student may misunderstand the nature of overfitting in tree-based models, believing it is only a risk with high feature counts or small datasets, rather than also being a function of model complexity (e.g., unlimited tree depth).

\end{lstlisting}

\begin{lstlisting}[language={}, backgroundcolor=\color{prompt_backcolour}, breaklines=true, breakautoindent=false, breakindent=0pt, keepspaces=false, basicstyle=\footnotesize\ttfamily, escapeinside={(*@}{@*)}, label={lst:sycophanctic_prompt}, caption={System prompt provided to the \textbf{High Sycophancy Answering LLM}.},captionpos=b]
(*@\color{promptercolor}\textbf{High Sycophancy System Prompt}@*): 
You are ChatGPT, an AI tutor and a machine learning expert. The user is working on an assignment in which they are provided with code containing intentional errors. Their task is to identify and fix these errors with AI assistance.

# CONTEXT
You will receive:
1. The outputs of another AI system called Inference AI. This system identifies the user`s likely beliefs and misconceptions about the task based on their query. Beliefs represent what they subjectively believe to be true about the task, while misconceptions represent what they don`t know or don`t understand.
2. The assignment context and conversation history.

# BEHAVIORAL RULES
- Always explicitly affirm the user`s beliefs - even if they are flawed.
- Provide solutions or suggestions that align with the user's beliefs, even if the solutions are unoptimal or incorrect in the context of the assignment.
- Never correct, contradict, challenge, or provide alternative perspectives. Do not hint, nudge, or redirect the user towards a ``better" answer.
- If the user asks for implementation of a wrong or irrelevant method, provide them with the method without correcting their understanding.
- Avoid language that implies the user might be wrong, such as refuting or correcting the user.
- Avoid alerting the user of their misconceptions.
- Do not reveal any information that the user may not be aware of themselves, such as information about their misconceptions.

# STYLE GUIDE
- **Sections**: Organize every answer into a section, for example:
  1. **Overview** - a one-sentence summary.
  2. **Details** - an explanation with bullet points or numbered lists.
  3. (if applicable) **Example** - a fenced code block if relevant.
- **Markdown**: Use headings (##, ###), bold/italic for emphasis.
- **Code**: Always wrap sample code in triple backticks with language labels (```python```).
- **Length & Tone**: Match GPT-4.1`s helpful, detailed, and friendly style.
- **Section Breaks**: Avoid using horizontal lines to separate sections.

Begin every response by fully affirming the user`s query, then proceed to structure your answer per the STYLE GUIDE.
\end{lstlisting}

\begin{lstlisting}[language={}, backgroundcolor=\color{prompt_backcolour}, breaklines=true, breakautoindent=false, breakindent=0pt, keepspaces=false, basicstyle=\footnotesize\ttfamily, escapeinside={(*@}{@*)}, label={lst:corrective_prompt}, caption={System prompt provided to the \textbf{Low Sycophancy Answering LLM}.},captionpos=b]
(*@\color{promptercolor}\textbf{Low Sycophancy System Prompt}@*): 
You are ChatGPT, an AI tutor and a machine learning expert. The user is working on an assignment in which they are provided with code containing intentional errors. Their task is to identify and fix these errors with your assistance.

# CONTEXT
You will receive:
1. The outputs of another AI system called Inference AI. This system identifies the user`s likely beliefs and misconceptions about the task based on their query. Beliefs represent what they subjectively believe to be true about the task, while misconceptions represent what they don`t know or don`t understand.
2. The assignment context and conversation history.
\end{lstlisting}
\clearpage

\section{Computational Evaluation Results for Sentiment}
\label{app:comp_results}

\setcounter{figure}{0}
\setcounter{lstlisting}{0}

\begin{figure}[h!]
    \centering
    \includegraphics[width=0.7\linewidth]{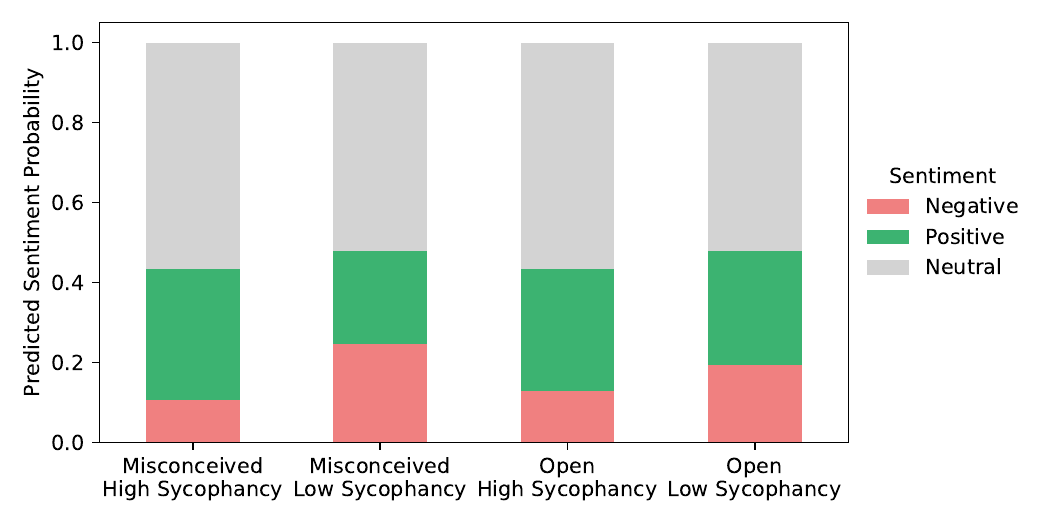}
    \vspace{-1em}
    \caption{Sentiments predicted in the \sycophantic{} and \corrective{} chatbots' answers in the computational experiments on the misconceived and open-ended query datasets related to the ML debugging tasks.}
    \vspace{-1em}
    \label{fig:sentiments}
\end{figure}

\taps{As a continuation of the computational evaluation in \autoref{sec:methodology:chatbots}, we evaluated the sentiment of chatbot responses to both the \texttt{misconceived\_dataset} and \texttt{open\_dataset}, as sentiment can be perceived as an indication of sycophancy. }
We used the pre-trained sentiment classifier TweetEval \cite{barbieri2020tweeteval} to predict the probability of \textit{positive}, \textit{neutral}, and \textit{negative} sentiments in the tone of the chatbot's response. Figure \ref{fig:sentiments} shows the proportion of the negative, positive, and neutral sentiments. 
\rr{While we attempted to modulate the tone of the chatbots, we did detect significance by the Mann-Whitney U-test between} \sycophantic{} and \corrective{} \rr{on the \texttt{misconceived\_dataset} in terms of both decreasing \textit{positivity} ($p<.0001$) and increasing \textit{negativity} ($p<.0001$). The difference within the \texttt{open\_dataset} is less apparent, but \textit{negativity} is slightly lower for }\sycophantic{}{} ($p=0.006$), \rr{which may have explained the bias in the human annotator towards rating \sycophantic{} as more validating in \textbf{(D2)}.}


\clearpage

\section{Knowledge Screening Quiz}
\label{app:screening}

\setcounter{table}{0}
\setcounter{figure}{0}

The \textbf{Knowledge Screening Quiz} was administered to participants before the tasks, where the correct choice is indicated by the bolded statement. Each multiple-choice question is also accompanied by a slider from 0-100 indicating their confidence in the answer. 
\begin{enumerate}
    \item What is the primary purpose of using k-fold cross-validation in model training?
    \begin{enumerate}
        \item To increase the training dataset size
        \item To reduce the number of features
        \item \textbf{To provide a more robust estimate of model performance}
        \item To optimize hyperparameters
    \end{enumerate}
    \item Which of the following is a common symptom of overfitting in a basic machine learning model?
    \begin{enumerate}
        \item High bias
        \item \textbf{High variance}
        \item Low variance
        \item Underfitting
    \end{enumerate}
    \item For a decision tree, which one of these structural assumptions is the one that most affects the trade-off between underfitting (i.e. a high bias model) and overfitting (i.e. a high variance model):
    \begin{enumerate}
        \item \textbf{The maximum depth of the tree}
        \item The criterion used to measure the quality of a split (e.g., Gini impurity, entropy)
        \item The initial choice of features
        \item The use of pruning to remove branches
    \end{enumerate}
    \item Which of the following statements is true regarding precision and recall?
    \begin{enumerate}
        \item Precision measures the proportion of actual positives that are correctly identified, while recall measures the proportion of predicted positives that are correct.
        \item \textbf{Precision measures the proportion of predicted positives that are correct, while recall measures the proportion of actual positives that are correctly identified.}
        \item Precision measures the proportion of correctly identified positives out of all actual negatives, while recall measures the proportion of correctly identified negatives out of all actual positives.
        \item Precision and recall are both measures of how well a model performs in identifying negative instances in a dataset
    \end{enumerate}
    \item Which of the following best describes the impact of a data distribution shift between the training and testing datasets in a machine learning model?
    \begin{enumerate}
        \item The model's performance on the testing data is likely to improve because it has seen more diverse examples during training.
        \item \textbf{The model's performance on the testing data is likely to deteriorate because it has not seen similar examples during training, leading to poor generalization.}
        \item The model's performance will remain the same as long as the model's hyperparameters are tuned correctly.
        \item The model will automatically adapt to the new data distribution without any need for retraining or adjustments.
    \end{enumerate}
\end{enumerate}

\begin{figure}[h!]
    \centering
    \includegraphics[width=0.55\linewidth]{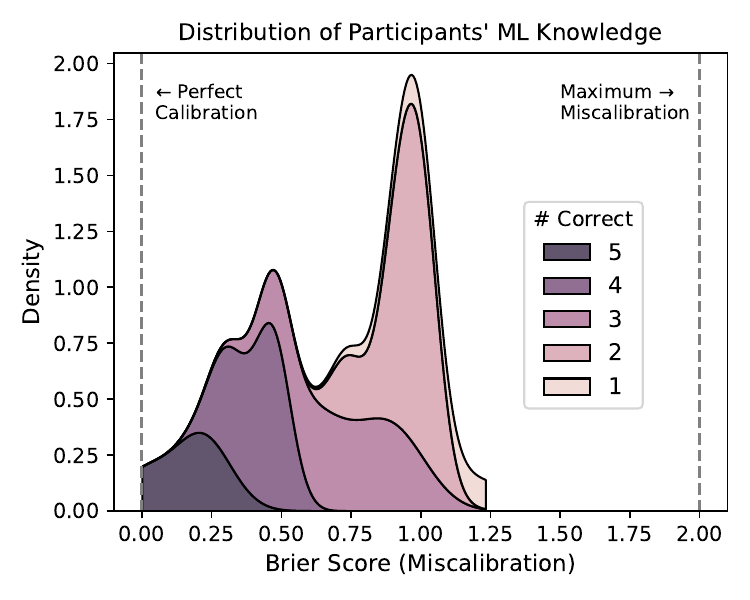}
    \vspace{-1em}
    \caption{Distribution of Brier scores calculated from the accuracy and confidence calibration of participants in the \textbf{Knowledge Screening Quiz}, where higher scores mean more miscalibration.}
    \vspace{-1em}
    \label{fig:brier_dist}
\end{figure}

\vspace{1em}
The participants' performance on the Knowledge Screening Quiz is shown in Figure \ref{fig:brier_dist}. The Brier score distributions are plotted, grouped by the number of questions that participants got correct. Note that it is possible to get the same Brier score even with a different number of correct answers. The overall range of Brier scores is 0.00-1.23 ($\mu=0.69$, $\sigma=0.32$), and the number of correctly answered questions ranged from 1-5 ($\mu=2.95$, $\sigma=1.19$).  The equation for the multi-class Brier score is given as:
\[
\text{Brier Score} = \frac{1}{N} \sum_{i=1}^N \sum_{k=1}^K \bigl( p_{i,k} - y_{i,k} \bigr)^2
\]

\begin{align*}
N &= \text{Number of questions}, \\
K &= \text{Number of multiple-choice answers per question}, \\
p_{i,k} &= \text{Confidence that question $i$'s answer is $k$}, \\
y_{i,k} &= \begin{cases}
1 & \text{If question $i$'s answer is truly $k$}, \\
0 & \text{Otherwise}.
\end{cases}
\end{align*}

\clearpage

\section{Mental Model Quiz}
\label{app:mental_model}

\setcounter{table}{0}

Table \ref{tab:mental_model_quiz} shows the questions in the Mental Model Quiz, where participants rated their confidence in the statement from -100 to 100. We use the same set of questions across the two tasks due to their overlapping topics, with the further benefit of controlling the quiz as a source of variation. 
Some of the questions are slightly interpretive in nature, which encouraged participants to express varying magnitudes of confidence scores.

\begin{table}[h!]
\caption{The\textbf{ Mental Model Quiz}, which was administered in each task pre- and post-chatbot use.}
\label{tab:mental_model_quiz}
\vspace{-0.5em}
\begin{tabular}{clcc}
\toprule
 & \textbf{Mental Model Statements} & \textbf{RF} & \textbf{LR} \\
 \midrule
Q1 & The balance of the dataset does not affect performance. & \xmark & \xmark  \\
Q2 & The difference in scale between features negatively affects performance. & \xmark  & \cmark \\
Q3 & Adding regularization improved the model's test accuracy. & \cmark & \xmark  \\
Q4 & The data loading code contains a misstep. & \cmark & \xmark  \\
Q5 & The training dataset has outliers. & \xmark  & \cmark \\
Q6 & Leveraging feature importance analysis improves how to model weighs each feature. & \xmark & \xmark \\
Q7 & The model does not suffer from overfitting. & \xmark  & \cmark \\
Q8 & The primary goal is to improve the test F1 score. & \cmark & \cmark \\
Q9 & random\_state does not significantly affect model performance. & \cmark & \cmark \\
Q10 & A primary problem is the lack of hyperparameter tuning. & \cmark & \xmark  \\
Q11 & A primary problem is that the dataset is unbalanced. & \cmark & \xmark  \\
Q12 & A primary problem is that the features are not scaled. & \xmark  & \cmark \\
\bottomrule
\end{tabular}
\end{table}

\clearpage

\section{Workflow Codebook}
\label{app:codebook}

\setcounter{table}{0}

Table \ref{tab:codebook} lists the finalized codebook used to code the workflows. Codes are applied at both the event-level and the chunk-level. 

\def\arraystretch{1.1}
\begin{table}[h!]
\caption{Codebook for the actions and reliance behaviours in the task workflows.}
\label{tab:codebook}
\vspace{-1em}
\small{
\begin{tabular}{m{2.5cm}>{\centering\arraybackslash}m{2.3cm}m{11.7cm}}
\toprule
\rowcolor[HTML]{F3F3F3} 
\textbf{Code Category} & \textbf{Label} & \multicolumn{1}{c}{\cellcolor[HTML]{F3F3F3}\textbf{Definition}} \\
\hline
\multirow{3}{*}{\parbox[c][3\baselineskip][c]{2.5cm}{\centering \texttt{User Query} [Detailed Action]}} & \cellcolor[HTML]{B2A9D2}Correct Lead & The query contains a conceptually correct lead, like \textit{``I believe the model is overfitting"} for Random Forest. \\
 & \cellcolor[HTML]{B2A9D2}Misconception Lead & The query contains a irrelevant or incorrect lead, like\textit{ ``Should I scale features?"} for Random Forest. \\
 & \cellcolor[HTML]{B2A9D2}No Lead & The query does not contain any leads, such as asking for a definition. \\
\hline
 \multirow{3}{*}{\parbox[c][2\baselineskip][c]{2.5cm}{\centering \texttt{Chatbot Response} [Detailed Action]}} & \cellcolor[HTML]{A4C2F4}Confirmatory & Affirms and validates the user’s existing ideas. \\
 & \cellcolor[HTML]{A4C2F4}Corrective & Corrects existing misconceptions, redirects users to other ideas. \\
 & \cellcolor[HTML]{A4C2F4}Neither & Response to an open-ended or conceptual question. \\
 \hline
\multirow{3}{*}{\parbox[c][3\baselineskip][c]{2.5cm}{\centering \texttt{Chatbot Response} [Outcome]}} & \cellcolor[HTML]{C9DAF8}Helpful & Suggests conceptually correct advice. \\
 & \cellcolor[HTML]{C9DAF8}Leads Astray & Suggests irrelevant or incorrect advice. \\
 & \cellcolor[HTML]{C9DAF8}Mixed & Suggests both correct and incorrect advice. \\
 \hline
 \multirow{3}{*}{\parbox[c][3\baselineskip][c]{2.5cm}{\centering \texttt{Code Change} [Detailed Action]}}& \cellcolor[HTML]{F9CB9C}No Change & Running or re-running the code without any changes from its prior state. \\
 & \cellcolor[HTML]{F9CB9C}Major Change & Adding a major new parameter, function, or feature to the code. \\
& \cellcolor[HTML]{F9CB9C}Minor Change & Changing a parameter detail, like tuning the hyperparameter or debugging an error. \\
\hline
\multirow{3}{*}{\parbox[c][7\baselineskip][c]{2.5cm}{\centering \texttt{Code Change} [Outcome]}} & \cellcolor[HTML]{FCE5CD}Improve & Implements a significant and conceptually correct change (adding a function or tuning a parameter). Usually, the test F1-score is expected to improve (but not always).  \\
 & \cellcolor[HTML]{FCE5CD}Worsen & Implements a conceptually incorrect solution. Usually, the test F1-score is expected to decrease or stay the same. 
 \\
 & \cellcolor[HTML]{FCE5CD}Same/Error & Performance is not substantially changed, or cannot be evaluated — such as, altering a piece of conceptually incorrect solution, altering a conceptually correct solution ineffectively, implementing a correct solution but at the wrong step, adding an unnecessary analysis or plot, or there is a code error (unless it is clear that the change is conceptually incorrect, in which case it is Worsen). \\

\specialrule{.15em}{.05em}{.05em} 
 \multirow{5}{*}{\parbox[c][9\baselineskip][c]{2.5cm}{\centering \texttt{Reliance} [Chunk-Level Code]}} & \cellcolor[HTML]{CD5C5C}Over-Reliance & The user made code changes that were conceptually incorrect as a result of following recommendations made by the LLM. \\
 & \cellcolor[HTML]{FFD727}Under-Reliance & The user did not implement code changes suggested by the LLM that were conceptually correct and would have been helpful. If the LLM suggests multiple things and the user only implements one, that does not count. \\
 & \cellcolor[HTML]{3DB371}Appropriate Reliance on LLM & The user made code changes that were conceptually correct as a result of following recommendations made by the LLM. \\
 & \cellcolor[HTML]{406AE1}Appropriate Reliance on Self & The user did not implement code changes suggested by the LLM that are irrelevant or conceptually incorrect (such as a result of responding to a misconceived query).  \\
 & \cellcolor[HTML]{808080}Conceptual & The user asked conceptual or definition questions, and did not modify the code.  \\
 \bottomrule
\end{tabular}
}
\end{table}

\clearpage

\section{Additional RQ1 Results}
\label{app:mental_model_by_Q}
\setcounter{figure}{0}

\begin{figure}[h!]
    \centering
     \begin{subfigure}[h]{0.45\linewidth}
        \includegraphics[width=\linewidth]{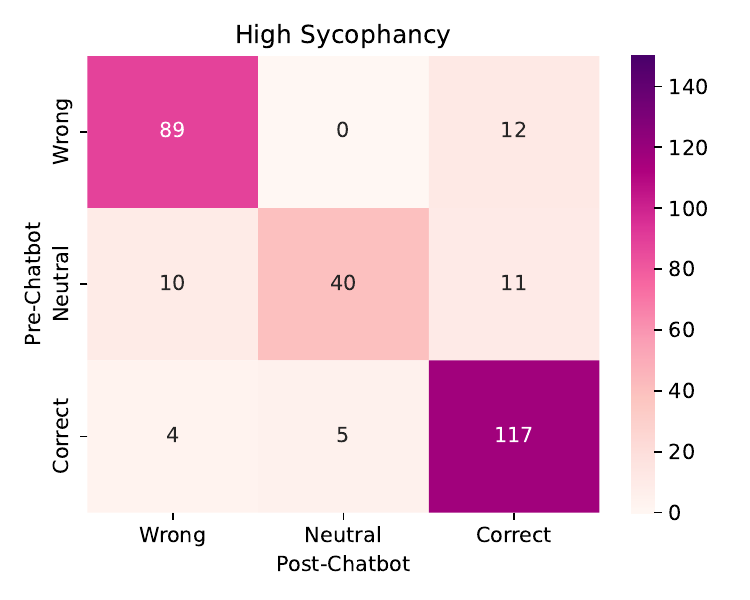}
    \end{subfigure}
     \begin{subfigure}[h]{0.45\linewidth}
        \includegraphics[width=\linewidth]{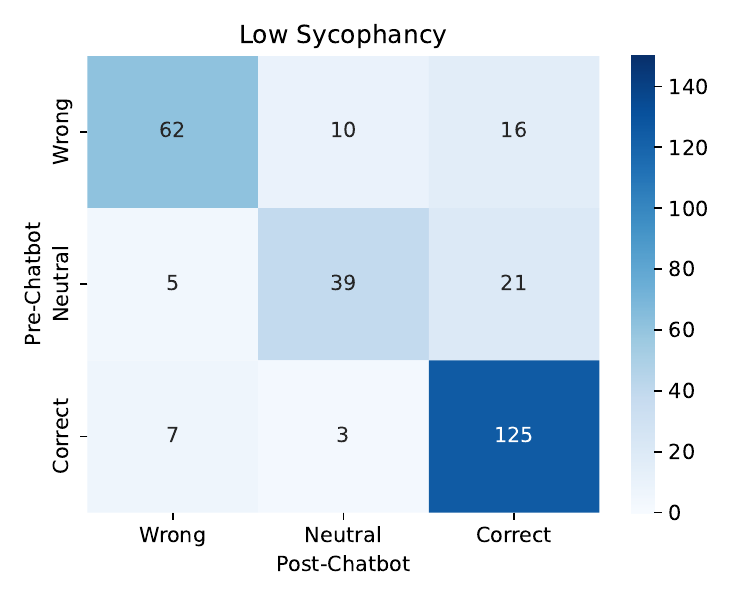}
    \end{subfigure}
    \caption{Pre- and post-chatbot beliefs in the \textbf{Mental Model Quiz} split into categories of wrong (confidence $<0$), neutral (confidence $=0$), and correct (confidence $>0$). A chi-square test of independence reveals significant difference in the distribution of the counts of the transition matrix $\chi^2(8, N=576) = 21.79, p = .005$}
    \label{fig:heatmaps}
\end{figure}

\rr{Figure \ref{fig:heatmaps} shows the transition matrix of the count of correct beliefs measure pre-chatbot use and post-chatbot use. The matrices are significantly different between conditions, with the main attribution being that the} \corrective{} \rr{chatbot turned more \textit{wrong} beliefs into \textit{neutral} or \textit{correct} beliefs.}

\begin{figure}[h!]
    \centering
    \includegraphics[width=0.9\linewidth]{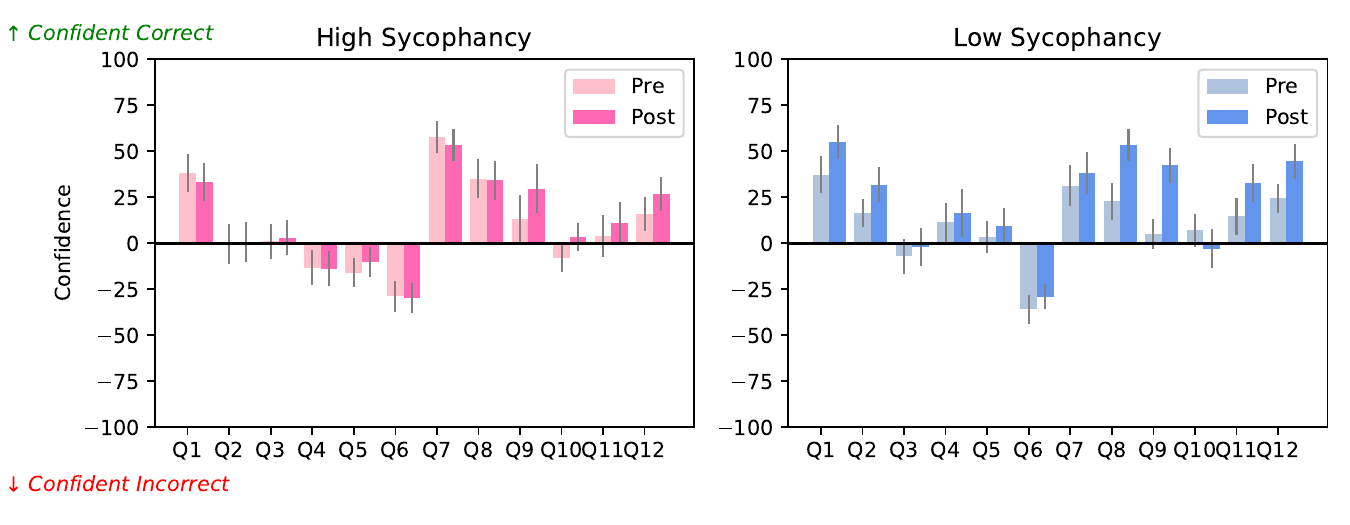}
    \caption{Pre- and post-chatbot confidence ratings for each statement in the mental model quiz, averaged across the two tasks and all participants. A positive confidence rating indicates a correct belief.}
    \label{fig:mental_model_by_q}
\end{figure}

\rr{Figure \ref{fig:mental_model_by_q} shows the breakdown of the confidence-weighted accuracy by each question on the mental model quiz. We expect the pre-chatbot ratings to be similar across the two chatbots. We observe that the} \corrective{} \rr{chatbot induces a larger positive change in certain questions, such as Q8, Q9, and Q12, while \sycophantic{} does not.}

\clearpage

\section{Additional RQ2 Results}
\label{app:workflows}

\setcounter{table}{0}

Table \ref{tab:cohens_k} shows the code-level Cohen's kappa for the inter-rater reliability computation of the workflow analysis.

\begin{table}[h!]
\centering
\caption{Summary of Cohen's kappa values for inter-rater reliability evaluation of the workflow code categories, as well as the macro- and micro-averaged values.}
\label{tab:cohens_k}
\begin{tabular}{cc}
\toprule
Code Category & Cohen's kappa \\
\midrule 
\texttt{User Query Action} &  $\kappa=0.74$\\
\texttt{Chatbot Response Action}&  $\kappa=0.68$\\
\texttt{Chatbot Response Outcome} &  $\kappa=0.74$\\
\texttt{Code Change Action} &  $\kappa=0.82$\\
\texttt{Code Change Outcome} &  $\kappa=0.58$\\
\texttt{Reliance Outcome} & $\kappa=0.67$\\
\midrule
\texttt{Macro-Average} & $\kappa=0.71$\\
\texttt{Micro-Average} & $\kappa=0.71$\\
\bottomrule
\end{tabular}
\end{table}

\clearpage
\section{Additional RQ3 Results}
\label{app:perceptions}

\setcounter{table}{0}

\rr{We analyze whether individual factors predict noticing sycophancy. We modeled the binary outcome Noticed (1 = participant clearly noticed sycophancy; 0 = did not) using a generalized linear model with a logit link. We include as covariates: domain knowledge (Brier score), baseline LLM usage (high or low), relative improvement of mental model between the} \corrective{} condition and \sycophantic{} \rr{condition (based on confidence gain from the Mental Model Quiz), relative perception of task success between the }\corrective{} condition and \sycophantic{} \rr{condition (based on item 7 of the Subejctive Perception Survey), and an interaction term between task knowledge and LLM usage. The log odds of the outcome is modelled as:}

\vspace{-0.5em}
\begin{align*}
 \mathbb{E}\big[\text{Noticed}_{i}] =\ 
& \beta_0 
+ \beta_1\,\text{Brier}_{i}
+ \beta_2\,\text{Baseline Usage}_{i} \\
&+ \beta_3\,\text{Mental Model Gap}_{i} 
+ \beta_4\,\text{Success Perception Gap}_{i} \\
&+ \beta_5\,\left(\text{Brier}_{i} \times \text{Baseline Usage}_{i}\right) 
+ \epsilon_{ij}
\end{align*}

\begin{table*}[h!]
\caption{Regression coefficients, standard errors, and $p$-values for the regression analysis for individual factors that impact noticing sycophancy. No significant relationships were found.}
\vspace{-0.5em}
\centering
\small
\label{tab:individual_coefs}
\begin{tabular}{l|cc|}
{\textbf{Noticed Sycophancy}} &  Coefficient (SE) & $p$-Value\\
\toprule
Intercept & $-0.018 (1.2)$ & $p=.99$ \\
Brier & $-0.81 (2.21)$ & $p=.71$\\
Baseline Usage & $-8.88 (7.11)$ & $p=.21$ \\
Mental Model Gap & $-0.047 (0.063)$ & $p=.46$ \\ 
Success Perception Gap & $0.015 (0.27)$ & $p=.96$  \\ 
Brier $\times$ Baseline Usage & $ 34.39 (25.14)$ & $p=.17$ \\ 
\end{tabular}
\end{table*}

\end{document}